\documentclass[prx,twocolumn,superscriptaddress]{revtex4-2}
 \pdfoutput=1
\usepackage[utf8]{inputenc}
\usepackage{graphicx}
\usepackage[T1]{fontenc}
\usepackage{hyperref}
\usepackage{qcircuit}
\usepackage{braket}
\usepackage[version=3]{mhchem} 
\usepackage{here}
\usepackage{multirow}
\usepackage{amsfonts,amsmath,amssymb,amsthm}
\usepackage{bm}
\usepackage{comment}
\usepackage[normalem]{ulem}

\newtheorem{lemma}{Lemma}

\newcommand{\lstickx}[1]{\lstick{\makebox[1.5em][l]{$#1$}}}
\newcommand{\arrep}[1]{\ar @<4pt> @/^/[#1]|-{\mbox{ $\times L$ }}}
\newcommand{\measoz}[1] {\mbox{$\left< \hat{O}\right>$}}
\newcommand{\measham}[1] {\mbox{$\left< H\right>$}}
\DeclareMathOperator{\tr}{tr}

\usepackage{mathtools}
\theoremstyle{plain}
\newtheorem{thm}{Theorem}
\newtheorem*{thm*}{Theorem}

\begin{document}
\title{Sequential optimal selection of a single-qubit gate and its relation to barren plateau in parameterized quantum circuits}

\author{Kaito Wada}
\email{wkai1013keio840@keio.jp}
\affiliation{Department of Applied Physics and Physico-Informatics, Keio University, 3-14-1 Hiyoshi, Kohoku-ku, Yokohama, Kanagawa, 223-8522, Japan}

\author{Rudy Raymond}
\affiliation{IBM Quantum, IBM Japan 19-21 Nihonbashi Hakozaki-cho, Chuo-ku, Tokyo, 103-8510, Japan}
\affiliation{Quantum Computing Center, Keio University, 3-14-1 Hiyoshi, Kohoku-ku, Yokohama, Kanagawa, 223-8522, Japan}
\affiliation{Department of Computer Science, The University of Tokyo, 7-3-1, Hongo, Bunkyo-ku, Tokyo 113-0033, Japan}

\author{Yuki Sato}
\affiliation{Toyota Central R\&D Labs., Inc., Koraku Mori Building 10F, 1-4-14 Koraku, Bunkyo-ku, Tokyo 112-0004, Japan}
\affiliation{Quantum Computing Center, Keio University, 3-14-1 Hiyoshi, Kohoku-ku, Yokohama, Kanagawa, 223-8522, Japan}

\author{Hiroshi C. Watanabe}
\email{hcwatanabe@keio.jp}
\affiliation{Quantum Computing Center, Keio University, 3-14-1 Hiyoshi, Kohoku-ku, Yokohama, Kanagawa, 223-8522, Japan}

\begin{abstract}
We propose an algorithm for variational quantum algorithms (VQAs) to optimize the structure of parameterized quantum circuits (PQCs) efficiently.
The algorithm optimizes the PQC structure on-the-fly in VQA by sequentially replacing a single-qubit gate with the optimal one to minimize the objective function.
To directly find the optimal gate, our method uses the factorization of matrices whose elements are evaluated on a set of the slightly-modified circuits.
The matrix factorization enables us to not only unify the existing sequential methods for further extension but also provide rigorous proofs of their limitation and potential in comparison with conventional gradient-based optimizers.
Firstly, when the circuits are sufficiently deep, the sequential methods encounter a barren plateau that the spectrum of the matrix concentrates on a single value exponentially fast with respect to the number of qubits.
Secondly, if the objective functions are local observables, they can avoid barren plateaus as long as the depth of the $n$-qubit PQCs is $\mathcal{O}(\log{n})$.
Although the family of these optimizers does not directly employ gradients of the objective function, our results establish their connection with conventional optimizations providing a consistent picture of the barren plateau.
We also perform numerical experiments showing the advantages over conventional VQAs and confirm the successful optimization getting over the barren plateau in the ground state problem of the mixed field Ising model up to 12 qubits.
\end{abstract}

\maketitle

\section{Introduction} \label{sec:introduction}
Variational Quantum Algorithm (VQA) is a notable classical-quantum hybrid algorithm, which is feasible on near-term quantum devices~\cite{cerezo2021variational}. It is a versatile methodology applicable to quantum chemical calculation~\cite{Peruzzo2014NatCom, Kandala2017Nat,gao2021applications}, combinatorial optimization problem~\cite{fuller2021approximate, amaro2022case, zoufal2022variational}, dynamics simulation~\cite{li2017efficient, yao2021adaptive}, time evolving simulation~\cite{Benedetti_2021, wada2022}, principle component analysis~\cite{larose2019variational, cerezo2022variational}, linear~\cite{bravo2019variational, xu2021variational} and nonlinear~\cite{lubasch2020variational,demirdjian2022variational} system solvers, and so on.
The key component in VQAs is a parameterized quantum circuit (PQC), also called ansatz, which is a sequence of quantum gates with classically controllable parameters~\cite{cerezo2021variational,tilly2021}.

The objective/cost function of VQA is often formulated as the expectation value of Hamiltonian $H$ of a target system whose solution can be obtained from the eigensystem of $H$ computed with variational quantum eigensolver (VQE), e.g., for quantum chemical calculations using fermionic or spin Hamiltonian~\cite{Peruzzo2014NatCom}.
Appropriate designs of PQCs are essential to express the quantum states of interest. 
PQCs are classified into physics-based ansatz and heuristic-based ansatz. 
A physics-based ansatz, including a unitary coupled cluster~\cite{shen2017quantum} and Hamiltonian variational ansatz~\cite{wecker2015progress, Wiersema2020PRXQ}, can achieve efficient optimization by limiting the Hilbert space spanned by the ansatz to the neighborhood of the target states.
However, it is difficult to run on near-term devices because of its deep circuit depth.
In contrast, a heuristic-based ansatz puts more weight on the feasibility on near-term devices, which generally results in shallower circuits albeit with the uncertainty to express the target states.

A typical strategy to mitigate the uncertainty is by systematically increasing the circuit depth with adding layers of gates. 
However, increasing layers can cause the gradient of the cost function with regards to parameters of PQCs to exponentially vanish as the number of qubits grows, a phenomenon termed \textit{barren plateau}~\cite{McClean2018NatComm,wang2021}. 
The barren plateau renders gradient-based approaches useless. 
Several remedies, such as layerwise-learning~\cite{Skolik2021} and parameter correlation~\cite{volkoff2021}, have been proposed but they can only cope with noiseless conditions. 
The only effective strategy for the noise-induced barren plateau to date is reducing circuit depth.
Hence, there are two conflicting requirements of PQCs with heuristic ansatz; they should be deep enough to express target states but should be as shallow as possible to avoid both types of barren plateau.

A circuit structure optimization implemented in Rotoselect~\cite{ostaszewski2021} draws attention to deal this dilemma, which is shared in Variable Ansatz (VAns) algorithm~\cite{bilkis2021}.
Rotoselect is also one of sequential quantum optimizers such as NFT~\cite{nakanishi2020} (also termed Rotosolve in~\cite{ostaszewski2021}), where single-qubit gates are sequentially and analytically optimized without gradient.
It allows to select the optimal (single-qubit) rotational gate among $R_x(\theta)$, $R_y(\theta)$, and $R_z(\theta)$ to minimize the objective function with regards to a single-qubit gate in a PQC.
In this optimal selection, the method utilizes the periodicity of the objective function in $\theta$, as $R_x(\theta) = \cos{(\theta/2)} I - i \sin{(\theta/2)}X$ and similarly for $R_y$ and $R_z$.
Since Rotoselect is more flexible via choosing the rotation axes of the single-qubit gates through cost minimization, it is regarded as structural optimization of fixed-depth PQCs.
However, it has two drawbacks: the rotation axis is from a finite and discrete gate set, and each parameter of PQCs is updated locally. These make it prone to local optima.

To deal with the drawbacks, "Free-axis selection" (Fraxis) was proposed based on the representation of a single-qubit gate whose rotational axis is arbitrary 3-dimensional vector, but the rotational angle is fixed to $\pi$~\cite{watanabe2021}.
Fraxis is also a sequential quantum optimizer, but with more parameters (i.e., axes of rotations) to optimize simultaneously.
The optimal axis is obtained from a matrix diagonalization whose elements are computed from expectation values of ${H}$ for quantum states generated from PQCs, where the gate of interest is replaced by a set of unitaries.
Because Fraxis continuously varies the axis of each Fraxis gate $R_{\bm{n}}(\pi)$ in a PQC, it has higher degree of freedom and better expressibility with limited depth.
However, unlike the gradient-based optimizers, nothing is known about the properties of sequential optimizers with regards to barren plateaus:~\cite{tilly2021} suggested they might be inferior to the gradient-based ones despite their faster convergence.

\begin{figure}[tb]
 \centering
 \includegraphics[scale=0.25]{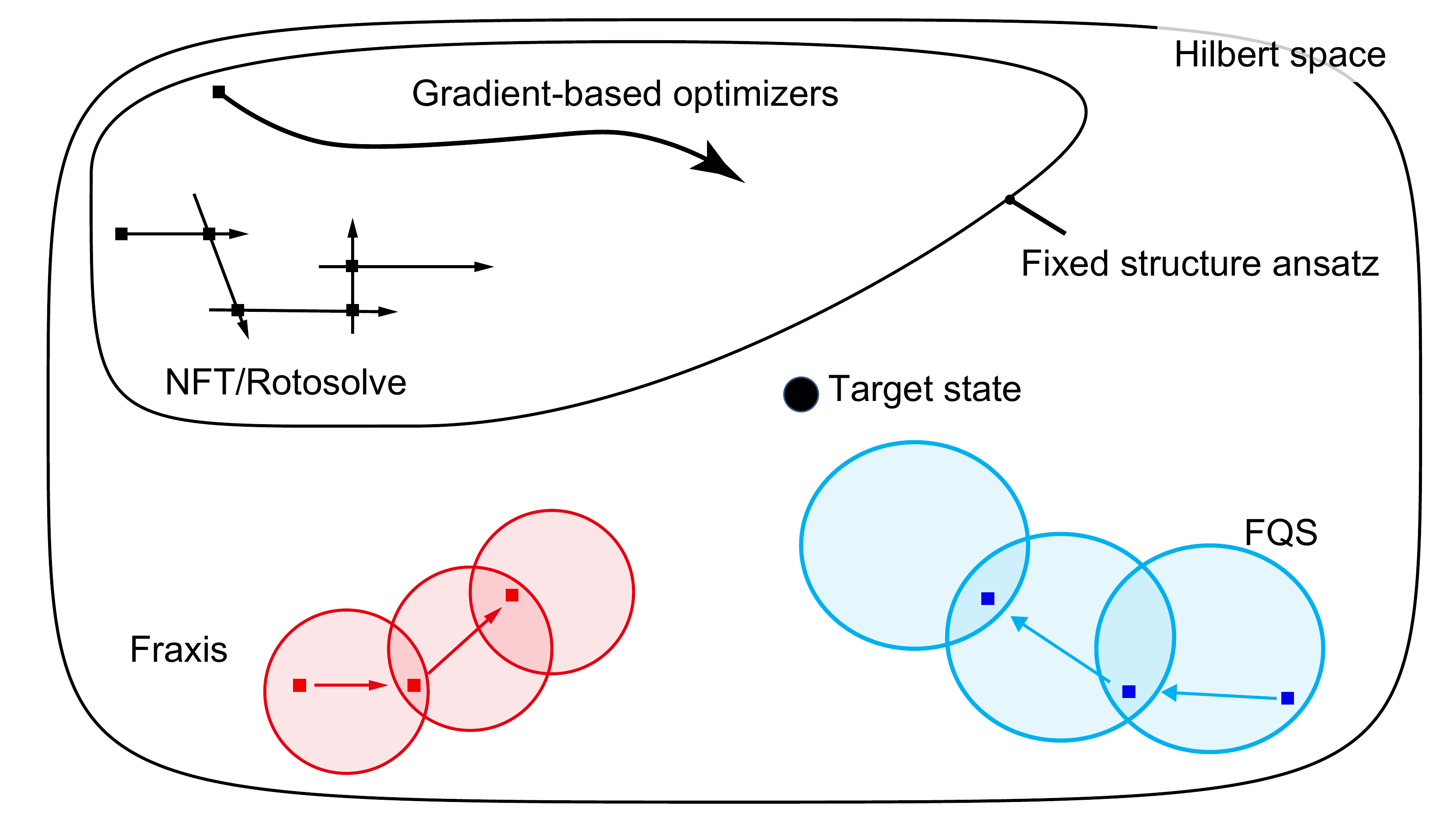}
 \caption{Schematic diagram of VQA with FQS in comparison with other optimizers. 
 The blue and red circles correspond to the expressibility of the single-qubit gate of FQS and Fraxis, respectively.
 Since FQS (Fraxis) has full (partial) degrees of freedom in a single-qubit gate, the FQS (blue) circles are larger than those of Fraxis (red).
 FQS and Fraxis can find the optimal state in its circle that drifts stepwise reflecting the circuit structure optimization.
 }\label{fig:scheme}
\end{figure}

In this work, we first extend the matrix diagonalization framework of Fraxis to full optimization of a general single-qubit gate, termed as \textit{Free Quaternion Selection} (FQS) after the quaternion representation of a single-qubit gate.
More precisely, to optimize one of single-qubit gates in a PQC, the VQA cost function can be transformed into a solvable quadratic form by mapping the single-qubit gate $R_{\bm{n}}(\psi)$ to a unit quaternion.
Because the single-qubit gates of PQCs with FQS have the highest degree of freedom, FQS can optimize more effectively the circuit structure to express target quantum states with limited depth as illustrated in Fig.~\ref{fig:scheme}.
In addition, we show that the FQS generalizes the existing sequential quantum optimizers; all of them can be formulated with the matrix diagonalization whose minimal eigenvalue corresponds to the (locally) optimal energy.

Next, we unveil several important properties of this optimizer family with respect to barren plateaus.
Originally, barren plateau was shown using the gradient-based framework where it manifests as exponential decay of the variance of gradient.
Since sequential quantum optimizers do not directly use gradients, it is not straightforward to study the properties of these optimizers with regards to barren plateaus.
However, the essential nature of barren plateaus is the loss of trainability as the number of qubits increases, which reflects an extreme separation between the initial and the target states caused by exponential inflation of the dimension of Hilbert space.
Hence, it is worth investigating the relation between loss of the trainability and the system size in sequential quantum optimizers.
To clarify this point, we introduce the spectral radius of the matrix associated with each optimizer as an alternative measure for trainability.
The spectral radius corresponds to the cost difference by a single application of sequential optimization to the gate of interest.
Therefore, if sequential optimizers are subject to the barren plateaus, the second moment of the spectral radius should exponentially vanish with respect to the system size.

Here, we rigorously prove that under the same conditions as assumed in their gradient-based counterparts~\cite{McClean2018NatComm,Cerezo2021NatComm}, the sequential optimizers will likely run into barren plateaus when a PQC has sufficient expressibility.
More precisely, if the circuits become sufficiently deep to achieve unitary 2-design over the whole system, the second moment of the spectral radius shows exponential decay regardless of the global or local cost.
On the other hand, we also rigorously prove that when the cost functions are local observables on an alternating layered ansatz~\cite{Cerezo2021NatComm,nakaji2021expressibility}, the exponential concentration of the spectrum cannot occur as long as the depth of the $n$-qubit PQCs is $\mathcal{O}(\log{n})$, and as a result the barren-plateau problems can be avoided.
The latter result is obtained thanks to the analysis of the cost-function dependent barren plateaus in~\cite{Cerezo2021NatComm} combined with our bounds on the spectral radius of the matrices used to optimize single-qubit gates sequentially.
We further demonstrate these properties using extensive numerical simulations up to 12 qubits.
Because the sequential optimizers have been shown faster in convergence, our results shed new light on sequential optimizers as better alternatives to the gradient-based ones.

The remainder of this paper is organized as follows.
In Sec.~\ref{sec:theory}, we describe the theoretical aspects of FQS.
We first give the quaternion representation for a single-qubit gate and then derive the matrix for FQS that completely characterize the energy landscape of the PQCs.
We then present the FQS as a generalized form of existing methods by showing their corresponding matrices.
The optimal energy can be determined from the optimal eigenvalues of the matrices. 
Rigorous proofs on the relation between barren plateaus with sequential optimizers are then derived.
In Sec.~\ref{sec:expetiments}, we provide extensive numerical experiments demonstrating the properties of FQS. 
Then, we confirm FQS outperforms other sequential quantum optimizers as well as gradient-based optimizers.
Moreover, in application to a mixed field Ising model up to 12 qubits, we demonstrate that VQE with FQS on an alternating layered ansatz for the local cost keeps the optimization capacity avoiding barren plateaus when the circuit is sufficiently shallow.
Finally, we conclude this study in Sec.~\ref{sec:conclusion}.

\section{Theory}\label{sec:theory}
\subsection{Quaternion representation for single-qubit gate}
A general single-qubit gate is conventionally represented as 
\begin{equation}\label{eq:general_1q_gate}
    R_{\bm{n}}(\psi):=\cos{\left(\frac{\psi}{2}\right)}I- i \sin{\left(\frac{\psi}{2}\right)}\bm{n}\cdot\vec{\sigma},
\end{equation}
where 
$I$ and $\vec{\sigma}=(\sigma_1,\sigma_2,\sigma_3)=(X,Y,Z)$ denote the $1$-qubit identity operator and the Pauli matrices.
The parameters $\bm{n}$ and $\psi$ correspond to a rotational axis and angle in the Bloch sphere, respectively.

Here, we show another way to parameterize the general single-qubit gate based on the well-known relationship between a single-qubit gate and a unit quaternion.
Since the rotational axis $\bm{n}$ is a three-dimensional real unit vector, we can write it in the polar coordinate system with the zenith angle $\theta$ and the azimuth angle $\phi$ as
\begin{equation}\label{eq:polar_S2}
    \bm{n}=\bm{n}(\theta,\phi)=(\cos{\theta},\sin{\theta}\cos{\phi},\sin{\theta}\sin{\phi}).
\end{equation}
Substituting Eq.~(\ref{eq:polar_S2}) into Eq.~(\ref{eq:general_1q_gate}), we obtain the quaternion representation of a single-qubit gate as 
\begin{equation}\label{eq:general_1q_gate_S3}
    R_{\bm{n}(\theta,\phi)}(\psi)=\bm{q}(\psi,\theta,\phi)\cdot\vec{\varsigma} \equiv R(\bm{q}),
\end{equation}
where a unit quaternion  $\bm{q}=(q_0,q_1,q_2,q_3)$ (i.e., $\bm{q}\in\mathbb{R}^4,~|\bm{q}|=1$) is parameterized with $(\psi,\theta,\phi)$ as
\begin{align}
\begin{split}
    q_0&=\cos{\left(\frac{\psi}{2}\right)},\\[4pt]
    q_1&=\sin{\left(\frac{\psi}{2}\right)}\cos{\theta},\\[4pt]
    q_2&=\sin{\left(\frac{\psi}{2}\right)}\sin{\theta}\cos{\phi},\\[4pt]
    q_3&=\sin{\left(\frac{\psi}{2}\right)}\sin{\theta}\sin{\phi}.
    \end{split}
\end{align}
Here, $\vec{\varsigma}=(\varsigma_0,\varsigma_1,\varsigma_2,\varsigma_3)$ is an extension of the Pauli matrices defined as
\begin{equation}
    \vec{\varsigma}:=\left(I,-iX,- iY,-iZ\right).
\end{equation}
Equation~(\ref{eq:general_1q_gate_S3}) allows us to identify a point $\bm{q}$ on the three-dimensional spherical surface with a single-qubit gate.
Based on this identification, we write $R_{\bm{n}}(\psi)$ as $R(\bm{q})$ for simplicity in the rest of this paper.  
Note that if we focus on the conventional single-qubit rotation gate with one parameter $\theta$ for the rotational angle, such as $R_{x}(\theta)$, this gate can be identified with a point on the one-dimensional spherical surface, i.e., the unit circle.

We can decompose a general single-qubit gate $R(\bm{q})$ into three $R_{z}$ gates and two $\sqrt{X}$ gates up to global phase~\cite{McKay_2017}, that is, 
\begin{align}
    R(\bm{q})= R_z(\phi)\sqrt{X}R_z(\theta)\sqrt{X}R_z(\lambda),
\end{align}
where $\theta,\phi,\lambda$ can be determined from $\bm{q}$.
Since $R_z$ gates on IBM Quantum devices are pulse-operation free, the general single-qubit gate can be implemented with only two times pulse-operation for $\sqrt{X}$.

\subsection{Our algorithm: Free Quaternion Selection for Variational Quantum Algorithm}

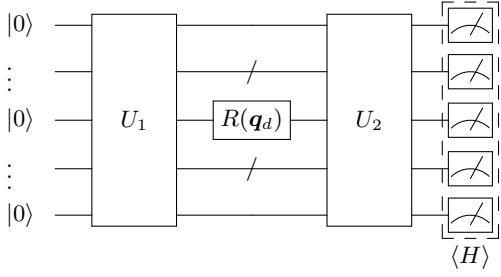
\begin{figure}[bht]
\centering
\begin{tabular}{c}
~~~~~\Qcircuit @C=1.5em @R=0.5em{
  \lstickx{\ket{0}} &\multigate{4}{~~U_1~~}
  &\qw 
  &\multigate{4}{~~U_2~~}
  &\meter \\
  \lstickx{\vdots} 
  &\ghost{~~U_1~~} 
  &{/} \qw  
  &\ghost{~~U_2~~} 
  &\meter 
  \\
  \lstickx{\ket{0}} 
  &\ghost{~~U_1~~}       
  &\gate{ R(\bm{q}_d) }
  &\ghost{~~U_2~~}       
  &\meter  \\
\lstickx{\vdots}
  &\ghost{~~U_1~~} 
  &{/} \qw  
  &\ghost{~~U_2~~} 
  &\meter 
  \\
  \lstickx{\ket{0}} 
  &\ghost{~~U_1~~}       
  &\qw     
  &\ghost{~~U_2~~}        
  &\meter
  \\ \\
  &&&&\measham{lll}\gategroup{1}{5}{5}{5}{.5em}{--}
}
\end{tabular}
\caption{
A quantum circuit to optimize the single-qubit gate $R(\bm{q}_d)$.
The variational parameters in $U_1$ and $U_2$ are fixed during the optimization of the gate of interest.
Our algorithm sequentially replace $R(\bm{q}_d)$
in the circuit by the optimal one, which is computed by the diagonalization of the corresponding matrix $S$ whose elements can be evaluated from the circuit by replacing $R(\bm{q}_d)$ with ten different gates as described in Sec.~\ref{subsec:evalS}.}
\label{fig:FQS_circuit}
\end{figure}
Let us consider an $n$-qubit parameterized quantum circuit $U$ consisting of $D$ parameterized single-qubit gates $\{R(\bm{q}_d)\}_{d=1}^D$ and parameter-free gates such as CNOT gate.
Tuning the parameters, we aim to solve an optimization task with the following objective function
\begin{align}
    \sum_{k=1}^K {\tr}\left[\rho_{k} U\left(\{\bm{q}_d\}_{d=1}^D\right)^\dagger H_k U{\left(\{\bm{q}_d\}_{d=1}^D\right)}\right],
\end{align}
where $\rho_k$ is an $n$-qubit initial state from a training set, and $H_k$ is some observable.
In the following, without loss of generality, we focus on a single expected value in the objective function (i.e., $K=1$, which can simply be considered as the minimization of energy for the Hamiltonian $H:=H_1$ and the input state $\rho_{\rm in}:=\rho_1$). 

The extension of our algorithm to the whole objective function is trivial due to the linearity.
For the energy minimization, we focus on a sequential optimization regarding $R(\bm{q}_d)$ where all parameters are fixed except for the $d$th single-qubit gate as shown in Fig.~\ref{fig:FQS_circuit}.

The energy expectation of the variational state is written as
\begin{align}\label{eq:obj_fn}
    \langle H\rangle(\bm{q}_d)
    &:={\tr}\left[\rho_{\rm in} U_1^\dagger R(\bm{q}_d)^\dagger U_2^\dagger  H U_2R(\bm{q}_d)U_1\right],\notag\\
    &={\tr}\left[\rho'_{\rm in} R(\bm{q}_d)^\dagger  H' R(\bm{q}_d)\right],
\end{align}
where $U_1$ and $U_2$ are the quantum circuits before and after $R(\bm{q}_d)$, respectively.
$H'$ and $\rho'_{\rm in}$ are defined as
\begin{align}
    H':=U_2^\dagger HU_2,~~~\rho_{\rm in}':=U_1\rho_{\rm in} U_1^\dagger.
\end{align}
Here we omit the subscript $d$ for simplicity.
Substituting 
Eq.~(\ref{eq:general_1q_gate_S3}) into 
Eq.~(\ref{eq:obj_fn}), we can obtain the following quadratic form 
\begin{align}\label{eq:quadratic_form}
    \langle H\rangle(\bm{q})=\bm{q}^\top S\bm{q},
\end{align}
where superscript $\top$ denotes a transpose operation, and $S=(S_{\mu\nu})$ is a $4\times 4$ real-symmetric matrix whose elements are defined as
\begin{align}
    S_{\mu\nu}:=\frac{1}{2}{\rm tr}\left[\rho'_{\rm in}\left(\varsigma^{\dagger}_{\mu}H'\varsigma_{\nu}+\varsigma^{\dagger}_{\nu}H'\varsigma_{\mu}\right)\right],
\end{align}
and more explicitly
\begin{widetext}
\begin{align}\label{eq:matrix_S}
    S=\begin{pmatrix}
    {\rm tr}(H'\rho'_{\rm in})
    &\frac{i}{2}{\rm tr}(H'[\rho'_{\rm in},\sigma_1])
    &\frac{i}{2}{\rm tr}(H'[\rho'_{\rm in},\sigma_2])
    &\frac{i}{2}{\rm tr}(H'[\rho'_{\rm in},\sigma_3])\\[5pt]
    \cdot&{\rm tr}(H'\sigma_1\rho'_{\rm in}\sigma_1)
    &\frac{1}{2}\left[{\rm tr}(H'\sigma_1\rho'_{\rm in}\sigma_2
    +H'\sigma_2\rho'_{\rm in}\sigma_1)\right]
    &\frac{1}{2}\left[{\rm tr}(H'\sigma_1\rho'_{\rm in}\sigma_3
    +H'\sigma_3\rho'_{\rm in}\sigma_1)\right]\\[5pt]
    \cdot&\cdot&{\rm tr}(H'\sigma_2\rho'_{\rm in}\sigma_2)
    &\frac{1}{2}\left[{\rm tr}(H'\sigma_2\rho'_{\rm in}\sigma_3
    +H'\sigma_3\rho'_{\rm in}\sigma_2)\right]\\[5pt]
    \cdot&\cdot&\cdot&{\rm tr}(H'\sigma_3\rho'_{\rm in}\sigma_3)
    \end{pmatrix},
\end{align}
\end{widetext}
where $[\cdot,\cdot]$ denotes the commutation relation.
See Appendix~\ref{apdx:A} for the derivation of the quadratic form.

The matrix $S$ can be obtained by running and measuring ten quantum circuits, each corresponding to the element in the upper diagonal of $S$, as detailed in the next subsection.
Note that the minimization of the quadratic form is exactly achieved by calculating the eigenvector corresponding to the lowest eigenvalue of $S$.
In addition, this optimization over the whole $SU(2)$ is a generalization of other sequential optimizers~\cite{nakanishi2020,ostaszewski2021,watanabe2021}, which optimize only a part of $SU(2)$.
To clarify this point, we show they can be derived from our general framework in Sec.~\ref{subsec:evalS}. 
Notice that a \textit{special} FQS, which applies to the objective functions in a special form, was proposed for time-evolving simulation~\cite{wada2022}.
Our formulation is also regarded as an extension of that special FQS.

Since FQS can select the optimal gate from the whole $SU(2)$ for minimizing the energy expectation in Eq.~(\ref{eq:obj_fn}), it can incorporate the multi-parameter correlation in $SU(2)$ and thus achieves better performance for optimizing PQCs than other sequential optimizers. 
Rotoselect, Rotosolve, and their variants~\cite{nakanishi2020,ostaszewski2021,VidalTheis2018,Parrish2019arXiv,koczor2022,Wierichs2022} can be used to optimize a general single-qubit gate by first decomposing the multi-parameter gate using ZYZ-decomposition~\cite{nielsen_chuang_2010} to obtain three single-parameter gates. 
In Rotosolve/select and their variants, each parameter of the decomposed gate is optimized locally in contrast to FQS and Fraxis, where  a general single-qubit gate is decomposed into two Fraxis gates defined as $R_{\bm{n}}(\pi)$. Fraxis~\cite{watanabe2021} updates these gates individually but simultaneously optimizes two parameters within the gate.

\subsection{Evaluation of the {\it S} elements}\label{subsec:evalS}

To determine the optimal single-qubit gate for the cost minimization, we evaluate the matrix $S$ constructing the quadratic form Eq.~(\ref{eq:quadratic_form}).
All the elements of $S$ in Eq.~(\ref{eq:matrix_S}) are calculated from ten expected values classified into three types as below. 
\begin{description}
  \item[Type-A]
  \begin{equation*}
      {\rm tr}\left[\rho'_{\rm in}H'\right]
  \end{equation*}
  \item[Type-B] for $k=1,2,3$,
  \begin{equation*}
  {\rm tr}\left[\rho'_{\rm in}\left( \frac{I\pm i\sigma_k}{\sqrt{2}}\right)^\dagger H'\left( \frac{I\pm  i\sigma_k}{\sqrt{2}}\right)\right]
  \end{equation*}
  \item[Type-C] for $(k,m)=(1,2),(1,3),(2,3)$
  \begin{equation*}
  {\rm tr}\left[\rho'_{\rm in}\left(\frac{\sigma_k+\sigma_m}{\sqrt{2}}\right)^\dagger H'\left(\frac{\sigma_k+\sigma_m}{\sqrt{2}}\right)\right].
  \end{equation*}
\end{description}
Here, Type-A is the $(0,0)$-the element of $S$ and corresponds to the expected value of $H$ on the PQC when the single-qubit gate of interest, as in Fig.~\ref{fig:FQS_circuit}, is replaced with identity.
The other diagonal elements are produced by the type-A and type-B values with the following identity:
\begin{align}
    &{\rm tr}(H'\sigma_k\rho'_{\rm in}\sigma_k)+{\rm tr}(H'\rho'_{\rm in})\notag\\
    &~~~={\rm tr}\left[\rho'_{\rm in}\left( \frac{I- i\sigma_k}{\sqrt{2}}\right)^\dagger H'\left( \frac{I- i\sigma_k}{\sqrt{2}}\right)\right]\notag\\
    &~~~~~~~~~+{\rm tr}\left[\rho'_{\rm in}\left( \frac{I+i\sigma_k}{\sqrt{2}}\right)^\dagger H'\left( \frac{I+ i\sigma_k}{\sqrt{2}}\right)\right].
\end{align}
Note that the Type-B corresponds to the expected value of $H$ on the PQC when the single-qubit gate of interest, as in Fig.~\ref{fig:FQS_circuit}, is replaced with, respectively, $R_x(-\pi/2)$, $R_x(\pi/2)$, $R_y(-\pi/2)$, $R_y(\pi/2)$, $R_z(-\pi/2)$, and $R_z(\pi/2)$.
In contrast, subtracting the type-B values with different signs yields the other elements in the first row directly.
The remaining off-diagonal elements are produced by the type-C expected values and the already obtained diagonal elements with the following identity:
\begin{align}
    &\frac{1}{2}\left[{\rm tr}(H'\sigma_i\rho'_{\rm in}\sigma_j)
    +{\rm tr}(H'\sigma_j\rho'_{\rm in}\sigma_i)\right]\notag \\
    &~~~= {\rm tr}\left[\rho'_{\rm in}\left(\frac{\sigma_i+\sigma_j}{\sqrt{2}}\right)^\dagger H'\left(\frac{\sigma_i+\sigma_j}{\sqrt{2}}\right)\right]\notag \\
    &~~~~~~~~~-\frac{1}{2} \left[ {\rm tr}(H'\sigma_i\rho'_{\rm in}\sigma_i) + {\rm tr}(H'\sigma_j\rho'_{\rm in}\sigma_j) \right].
\end{align}
Note that the Type-C values correspond to the expected values of $H$ on the PQC when the single-qubit gate to be optimized, as in Fig.~\ref{fig:FQS_circuit}, is replaced with, respectively, Fraxis gate $R_{\bm{n}}(\pi)$ for $\bm{n}\propto (1, 1, 0), (1, 0, 1), (0, 1, 1)$.
All expected values of Type-A, B, and C can be evaluated with direct measurements without any control operation such as the Hadamard test. 
Since the degree of freedom for a $4\times 4$ real-symmetric matrix is ten, the number of required direct measurements should be optimal.

\subsection{Unification of sequential quantum optimizers}
FQS generalizes all known sequential optimizers for single-qubit gates of PQCs, such as, Rotosolve (NFT), Rotoselect, Fraxis; those methods can be regarded as special cases of FQS.

For NFT, a single-qubit gate is restricted to a fixed axis $\bm{m}$ such as $U_{\rm NFT}:=R_{\bm{m}}(\psi)$.
Then the corresponding objective function in quadratic form is
\begin{align}\label{eq:quadratic_form_NFT}
    {\rm tr}\left[\rho'_{\rm in}U^\dagger_{\rm NFT}H'U_{\rm NFT}\right]
    =\boldsymbol{c}^\top\begin{pmatrix}
    S_{00}&\vec{S}_0\cdot\boldsymbol{m}\\    \vec{S}_0\cdot\boldsymbol{m}&\boldsymbol{m}^\top\tilde{S}\boldsymbol{m}\\
    \end{pmatrix}
    \boldsymbol{c},
\end{align}
where $\bm{c}:=(\cos{\psi/2},\sin{\psi/2})^\top$ and $\vec{S}_{0}:=(S_{01},S_{02},S_{03})$. $\tilde{S}$ denotes the lower right $3\times 3$ part of the $S$ matrix.
The derivation of the quadratic form is detailed in Appendix A.
The real symmetric matrix in Eq.~(\ref{eq:quadratic_form_NFT}) can be regarded as a contraction of the FQS matrix $S$ with respect to the rotational axis $\bm{m}$ reducing its degree of freedom to three.
In Rotoselect, three contracted matrices in Eq.~(\ref{eq:quadratic_form_NFT}) are constructed for $\bm{m} \in \{(1,0,0),(0,1,0),(0,0,1)\}$, and the lowest eigenvalue is selected after three separate diagonalization procedures.
Since the three matrices share $S_{00}$, the total number of circuit evaluations can be reduced to seven.

As for Fraxis, the target gate $U_{\rm Fraxis}:=R_{\bm{n}}(\pi)$ is simply expressed by the quaternion $q=(0,\bm{n})$.
Thus, substituting $q=(0,\bm{n})$ into the objective function, we can reproduce the previous results derived in \cite{watanabe2021} as follows
\begin{align}\label{eq:quadratic_form_Fraxis}
    {\rm tr}\left[\rho'_{\rm in}U^\dagger_{\rm Fraxis}H'U_{\rm Fraxis}\right]=\boldsymbol{n}^\top \tilde{S}\boldsymbol{n}.
\end{align}
Note that the $\theta$-Fraxis of~\cite{watanabe2021}, in which the rotation angle is fixed to arbitrary values $\theta$ instead of $\pi$, can be regarded as minimizing Eq.~(\ref{eq:quadratic_form}) for $\bm{q}':=(q_1,q_2,q_3)\in\mathbb{R}^3$ under the constraint $|\bm{q}'|^2=1-q_0^2$, which results in solving simultaneous equations rather than diagonalization.

It is worth noting that the required number of circuit evaluations for each sequential quantum optimizer coincides with the degrees of freedom for the real-symmetric matrix of respective methods, i.e.,  3=1+2 circuit evaluations for NFT/Rotosolve, 6=1+2+3 circuit evaluations for Fraxis, and 10=1+2+3+4 circuit evaluations for FQS.

\subsection{Barren plateaus in sequential quantum optimizers}
We have shown that the energy landscape of all sequential optimizers can be derived from the eigenvalues of the $S$ matrices whose elements are computed from circuit evaluations of slightly modified PQCs.
Sequential quantum optimizers can effectively obtain a better single-qubit gate minimizing the cost function as long as the eigenvalues of $S$ are \textit{significantly far} from degenerate.
Hence, it is useful to quantify degeneracy of $S$ for evaluation of performance of sequential quantum optimizers.
To this end, we introduce a centered matrix $S^{(p)}_{\rm c}$ defined as,
\begin{align}\label{eq:centered_S_matrix}
S^{(p)}_{\rm c} :=  S^{(p)}-\frac{{\rm tr}\left[S^{(p)}\right]}{p}I_{p\times p},
\end{align}
where $I_{p\times p}$ denotes a $p\times p$ identity matrix. Here, $S^{(p)}$ is a $p\times p$ real-symmetric matrix for each sequential quantum optimizer, and the degrees of freedom of target single-qubit gate is $p-1$ e.g., $p=2$ (Rotosolve/NFT), $p=3$ (Fraxis), and $p=4$ (FQS).
Henceforth, we omit the superscript $p$ for simplicity because it is clear from the context. 
Since the mean of the eigenvalues of $S_{\rm c}$ is zero, the spectral radius of $S_{\rm c}$ indicates the spread of the spectrum of $S$ from the mean of eigenvalues of $S$.
More precisely, defining $\hat{\lambda}(S)$ as the eigenvalue of $S$ with the largest distance from ${\rm tr}[S]/p$, which is the mean of eigenvalues of $S$, the spectral radius $r$ of $S_{\rm c}$ is equivalent to
\begin{align}\label{eq:spectralrad}
    r\left(S_{\rm c}\right) \equiv \left|\hat{\lambda}\left(S\right)-\frac{{\rm tr}\left[S\right]}{p}\right|.
\end{align}

In the following, we provide two theorems on the behavior of the spectrum of each real-symmetric matrix in Eqs.~(\ref{eq:quadratic_form}), (\ref{eq:quadratic_form_NFT}), and (\ref{eq:quadratic_form_Fraxis}) on the condition that the parameters in $U_1$ and $U_2$, as in Fig.~(\ref{fig:FQS_circuit}), are randomly initialized, which is standard for studying barren plateaus in gradient-based methods~\cite{McClean2018NatComm,Cerezo2021NatComm}.

\subsubsection{Sequential quantum optimizers catching barren plateaus}
An upper bound of Eq.~(\ref{eq:spectralrad}) that decays exponentially with the number of qubits hints at the existence of flat energy landscape (barren plateaus).
Indeed, we reveal that the flat landscape can happen under similar conditions assumed in~\cite{McClean2018NatComm} i.e., if the circuits become sufficiently deep to achieve unitary 2-design, the second moment of the spectral radius of the centered matrix shows exponential shrink regardless of the global or local cost.
\begin{thm}\label{theorem1}
    Suppose that the quantum circuits $U_1$ and $U_2$ are randomly and independently generated.
    If either $U_1$ or $U_2$ forms a unitary $t$-design with $t\geq 2$, the second moment of the spectral radius $r$ of the centered matrix Eq.~(\ref{eq:centered_S_matrix}) is upper bounded as,
    \begin{widetext}
    \begin{align}
        \mathbb{E}_{U_1,U_2}\left[r\left(S_{\rm c}\right)^2\right]\leq\begin{dcases}
        \frac{p^2{\rm tr}\left[H^2\right]\Delta \rho_{\rm in,2}}{2(2^{2n}-1)}+\frac{\Delta \rho_{\rm in,2}}{4p(2^{2n}-1)}\sum_{\mu,\nu=0}^{p-1}\left(p(-1)^{1-\delta_{\mu\nu}}-2\right){\rm tr}\left[\mathbb{E}_{U_2}\left[H'\varsigma_{\mu}\varsigma^\dagger_{\nu}H'\varsigma_{\nu}\varsigma^\dagger_{\mu}+{\rm h.c.}\right]\right]\\
        \frac{p^2{\rm tr}\left[\rho_{\rm in}^2\right]\Delta H_2}{2(2^{2n}-1)}+\frac{\Delta H_2}{4p(2^{2n}-1)}\sum_{\mu,\nu=0}^{p-1}\left(p(-1)^{1-\delta_{\mu\nu}}-2\right){\rm tr}\left[\mathbb{E}_{U_1}\left[\rho_{\rm in}' \varsigma^\dagger_{\mu}\varsigma_{\nu}\rho_{\rm in}' \varsigma^\dagger_{\nu}\varsigma_{\mu}+{\rm h.c.}\right]\right]
        \end{dcases}
        ,
    \end{align}
    \end{widetext}
    where the first case corresponds to $U_1$ being a $t$-design, and the second case corresponds to $U_2$ being a $t$-design.
    Here, $\mathbb{E}_{U_1,U_2}\left[\cdot\right]$ is defined as the expectation over the random quantum circuits $U_1$ and $U_2$,  $\delta_{\mu\nu}$ denotes the Kronecker delta, {\rm h.c.} means the Hermite conjugate of the preceding term, and $\Delta \rho_{{\rm in},2},~\Delta H_{2}$ are defined as
    \begin{align}
        \Delta \rho_{\rm in,2}&:={\rm tr}\left[ \rho_{\rm in}^2\right]-\frac{1}{2^{n}},\notag\\
        \Delta H_2&:={\rm tr}\left[ H^2\right]-\frac{{\rm tr}^2\left[ H\right]}{2^{n}}.
    \end{align}
\end{thm}
This theorem means that if either $U_1$ or $U_2$ has sufficient expressibility i.e., unitary $2$-design over $n$-qubit, the spectrum of $S$ matrix concentrates on a single value and its deviation is exponentially small with respect to the number of qubits.
As a result, this spectrum concentration implies that the energy of the output quantum state becomes insensitive exponentially on the selection of single-qubit gates.

\subsubsection{Sequential quantum optimizers avoiding barren plateaus}\label{subsubsec:al_bp}
Although Theorem~\ref{theorem1} assumes the relatively deep circuit forming unitary $t$-design with $t\geq 2$, we are interested in shallower circuits due to the existence of noise-induced barren plateau.
Here, we show that, in contrast to Theorem~\ref{theorem1}, there are shallow circuits avoiding the exponential shrink of the spectral radius when the target Hamiltonian is local.

\begin{figure}[thb]
   \centering
   \begin{tabular}{ccc}
   \begin{minipage}[c]{0.45\hsize}
        \centering
        \includegraphics[scale=0.27]{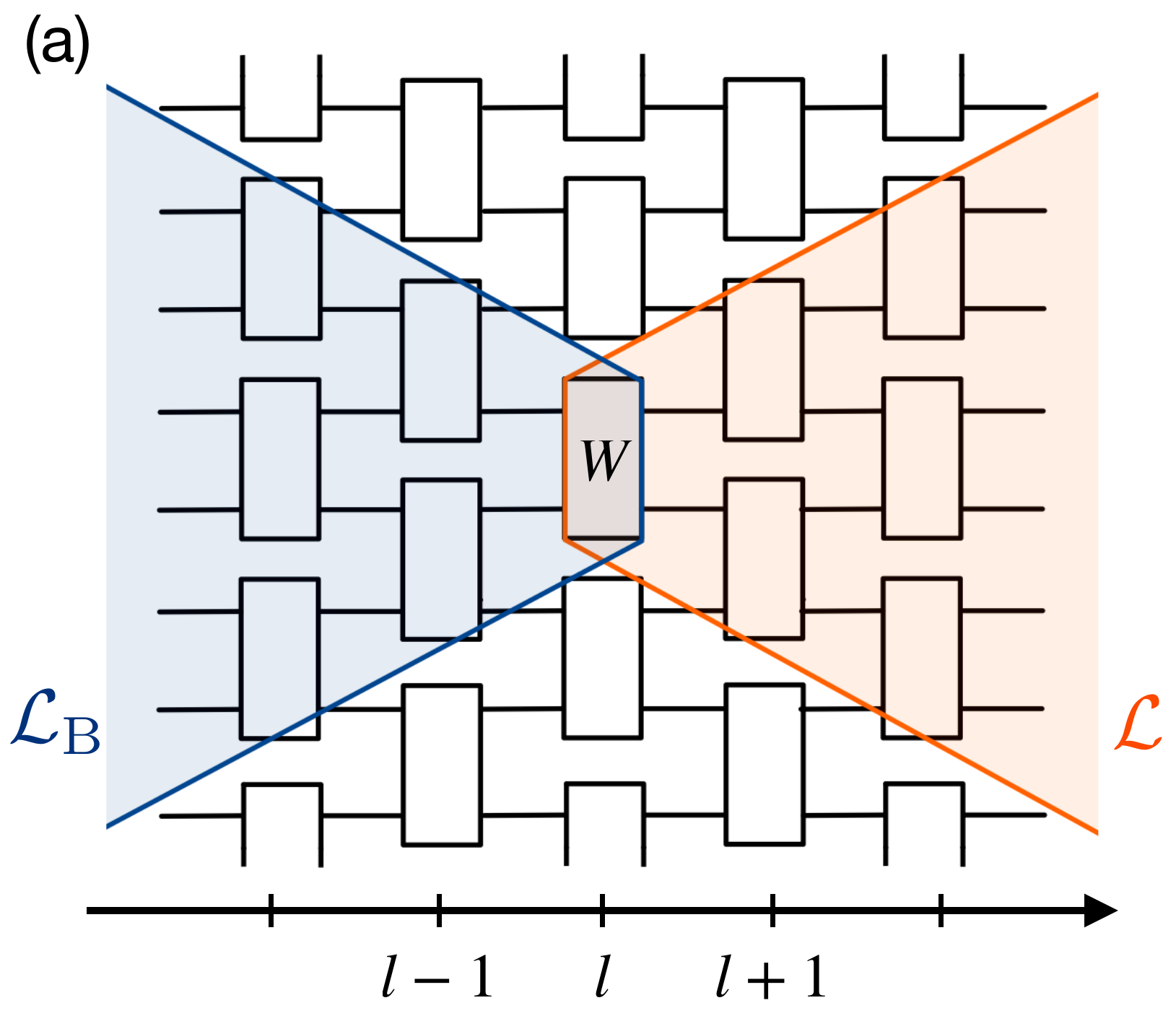}
   \end{minipage}
   &~~~&
   \begin{minipage}[c]{0.45\hsize}
        \centering
        \includegraphics[scale=0.3]{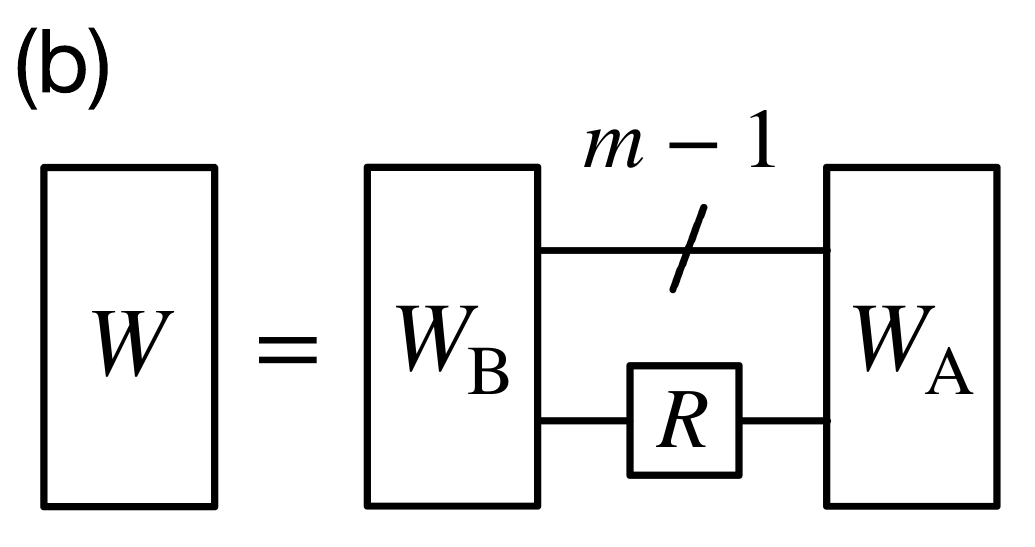}
   \end{minipage}
   \end{tabular}
   \caption{Schematic diagram of the alternating layered ansatz with $m$-qubit untiary blocks.
   In Theorem~\ref{theorem2}, we focus on a single-qubit gate $R$ in a block $W$ of the $l$th layer as depicted in (b), where $W_{\rm A}$ and $W_{\rm B}$ are a set of unitary gates located after and before the target single-qubit gate.
   $\mathcal{L}~(\mathcal{L}_{\rm B})$ denotes the forward (backward) light-cone with respect to $W$. Here, $U_1$ in Fig.~\ref{fig:FQS_circuit} includes the gates in $\mathcal{L}$ and $W_{\rm A}$, and $U_2$ includes the other gates.}
 \label{apdxfig:alansatz}
\end{figure}

To this end, let us consider an $n$-qubit alternating layered ansatz $U$~\cite{Cerezo2021NatComm,nakaji2021expressibility} with $m$-qubit parametrized unitary blocks, as depicted in Fig.~\ref{apdxfig:alansatz}(a).
Each block consists of parametrized single-qubit gates and parameter-free gates.
Suppose the ansatz consists of $L$ layers, and each layer contains $\xi$ blocks (i.e., $n=\xi m$).
We define $S_k~(k=1,2,\cdots,\xi)$ as the $m$-qubit subsystem on which the $k$th block from the top in the final layer acts.
In the following, we focus on a block $W$ in the $l$th layer and a parameterized single-qubit gate $R$ in the block as in Fig.~\ref{apdxfig:alansatz}(b).
Moreover, we define the forward light-cone $\mathcal{L}$ of the block $W$ as a series of gates that has at least one input qubit causally connected to the output qubits of $W$.
As well, the backward light-cone $\mathcal{L}_{\rm B}$ of $W$ is defined in the reverse direction of $\mathcal{L}$.

\begin{thm}\label{theorem2}
Suppose that the whole quantum circuit $U$ is an $n$-qubit alternating layered ansatz with $m$-qubit blocks as described in Fig.~\ref{apdxfig:alansatz}.
    Here, we focus on a block $W$ in $l$th layer and a parameterized single-qubit gate $R$ in the block.
    We assume that the quantum circuits $W_{\rm A},W_{\rm B}\subset W$, which are located after and before the target gate respectively, and the other blocks form a local $2$-design independently.
    In addition, the Hamiltonian $H$ is assumed to be $m$-local such as
    \begin{align}
        H=c_0I^{\otimes n}+\sum_{i}c_i{h}_i,~c_0,c_i\in\mathbb{R},~(\exists i,c_i\neq 0),
        \end{align}
    where $h_i$ is a tensor product of Pauli matrices that acts non-trivially on at most $m$-qubit.
    (The same assumption was used in~\cite{Cerezo2021NatComm}.) Then, the second moment of the spectral radius $r$ of the centered matrix Eq.~(\ref{eq:centered_S_matrix}) satisfies
    \begin{widetext}
    \begin{align}\label{eq:thorem2}
    \mathbb{E}_{U_1,U_2}\left[r\left(S_{\rm c}\right)^2\right]    &\geq \frac{(p+2)(p-1)2^{m(l+1)-1}}{p(2^{2m}-1)^2(2^{m}+1)^{L+l}} \sum_{i\in i_{\mathcal{L}}}\sum_{\substack{(k,k')\in k_{\mathcal{L}_{\rm B}}\\k'\geq k}} c_i^2\epsilon(\rho_{k,k'})\epsilon(h_i),
    \end{align}
    \end{widetext}
    where $L$ denotes the total number of layers.
    Here, $i_{\mathcal{L}}$ is the set of $i$ indices whose associated operators $h_i$ act on qubits in the forward light-cone $\mathcal{L}$ of $W$, and $k_{\mathcal{L}_{\rm B}}$ is the set of $k$ indices whose associated subsystems $S_k$ are in the backward light-cone $\mathcal{L}_{\rm B}$ of $W$.
    The quantum state $\rho_{k,k'}$ is the reduced density matrix of the input state $\rho_{\rm in}$ on $S_k S_{k+1}\cdots S_{k'}$, and the function $\epsilon(M)$ for a matrix $M$ is defined as $\epsilon(M)=D_{\rm HS}(M,{\rm tr}(M)\mathbf{1}/d_M)$ where $D_{\rm HS}$ is the Hilbert-Schmidt distance and $d_M$ is the dimension of the matrix $M$.
\end{thm}

This theorem means the lower bound of the second moment of the spectral radius does not vanish exponentially fast when we employ shallow $L$, which leads to similar properties as the gradient-based counterparts~\cite{Cerezo2021NatComm}. 
More precisely, if at least one term $c_i^2 \epsilon(\rho_{k,k'}) \epsilon(h_i)$ vanishes no faster than $\Omega(1/{\rm poly}(n))$, and if the number of layers $L$ is $\mathcal{O}({\rm log}(n))$, then
\begin{align}
\mathbb{E}_{U_1,U_2}\left[r\left(S_{\rm c}\right)^2\right]= \Omega\left(\frac{1}{{\rm poly}(n)}\right)
\end{align}
holds.
This guarantees trainability of sequential quantum optimizers in the beginning of optimization.
On the other hand, if at least one term $c_i^2\epsilon(\rho_{k,k'})\epsilon(h_i)$
vanishes no faster than $\Omega(1/2^{{\rm poly}({\rm log}(n))})$, and if the number of layer $L$ is $\mathcal{O}({\rm poly}({\rm log}(n))$ then,
\begin{align}
\mathbb{E}_{U_1,U_2}\left[r\left(S_{\rm c}\right)^2\right]= \Omega\left(\frac{1}{2^{{\rm poly}({\rm log}(n))}}\right)
\end{align}
holds.
This suggests a transition region of trainability where the spread of the spectrum decays faster than polynomial but slower than exponential.
Note that Cerezo et al.~\cite{Cerezo2021NatComm} showed a detailed analysis of variance of gradient with respect to angles of fixed-axis rotation gates.
In contrast, for a global cost function, the spectral radius is expected to shrink exponentially even on a constant-depth alternating layered ansatz in line with the gradient-based methods as~\cite{Cerezo2021NatComm}.

\section{Numerical Experiments} \label{sec:expetiments}
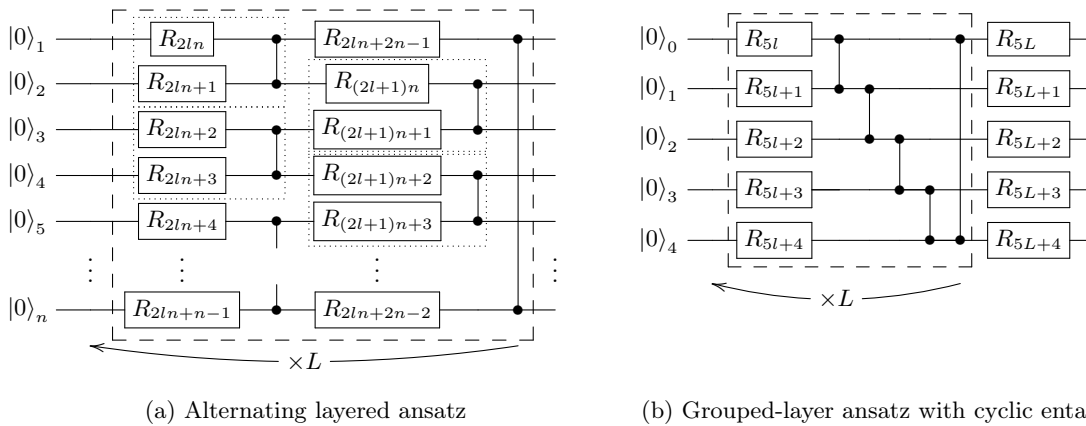
\begin{figure*}[t]
\centering
\begin{tabular}{ccc}

\Qcircuit @C=1.4em @R=.3em{
  \lstickx{\ket{0}_1}& \qw  & \gate{ R_{2ln}} & \ctrl{1}          & \gate{ R_{2ln+2n-1}} & \qw             & \control \qw &\qw       \\
  \lstickx{\ket{0}_2} & \qw & \gate{ R_{2ln+1}} & \control \qw  & \gate{ R_{(2l+1)n}} &\ctrl{1}          & \qw  &\qw        \\
  \lstickx{\ket{0}_3} & \qw & \gate{ R_{2ln+2}} & \ctrl{1}          & \gate{ R_{(2l+1)n+1}} & \control \qw & \qw    &\qw   \\
  \lstickx{\ket{0}_4} & \qw & \gate{ R_{2ln+3}} & \control \qw  & \gate{ R_{(2l+1)n+2}} & \ctrl{1}         &  \qw  &\qw  \\
  \lstickx{\ket{0}_5} & \qw & \gate{ R_{2ln+4}} & \ctrl{1}         & \gate{ R_{(2l+1)n+3}}  &\control \qw   & \qw   &\qw         \\
  &&            & &        &      & &  \\
  &&            & &        &      & &  \\
  &\vdots  &{\vdots}                            &             &{\vdots}                       & &&\vdots  \\
  &&            & &        &      & &  \\
  &&            & &        &      & &  \\
  &&            & &        &      & &  \\
  \lstickx{\ket{0}_n}     & \qw  & \gate{ R_{2ln+n-1}} & \ctrl{-1} &\gate{ R_{2ln+2n-2}}  & \qw              &\ctrl{-11} &\qw  \\
  &&&&&& \arrep{lllll}
  \gategroup{1}{3}{2}{4}{.4em}{..} 
  \gategroup{3}{3}{4}{4}{.4em}{..}  
  \gategroup{2}{5}{3}{6}{.4em}{..}  
  \gategroup{4}{5}{5}{6}{.4em}{..}  
  \gategroup{1}{3}{12}{7}{1.0em}{--}}
&~~~~~~~~~&
\Qcircuit @C=1.0em @R=.6em {
  \lstickx{\ket{0}_0} & \qw & \gate{R_{5l~~~}} & \ctrl{1}          & \qw              & \qw& \qw& \control\qw&\gate{R_{5L~~~}}    & \qw   \\
  \lstickx{\ket{0}_1} & \qw & \gate{R_{5l+1}} & \control \qw & \ctrl{1}          & \qw&\qw&  \qw&\gate{R_{5L+1}} & \qw  \\
  \lstickx{\ket{0}_2} & \qw & \gate{R_{5l+2}} & \qw        & \control \qw  & \ctrl{1}& \qw& \qw&\gate{R_{5L+2}}    & \qw  \\
  \lstickx{\ket{0}_3} & \qw & \gate{R_{5l+3}} & \qw \qw & \qw & \control \qw& \ctrl{1} \qw& \qw&\gate{R_{5L+3}}  & \qw   \\
  \lstickx{\ket{0}_4} & \qw & \gate{R_{5l+4}} & \qw              & \qw  & \qw& \control \qw& \ctrl{-4}\qw&\gate{R_{5L+4}}   & \qw  \\
  & & & & & &&\arrep{llllll}
  \gategroup{1}{3}{5}{8}{.7em}{--}}
\\
\\
(a) Alternating layered ansatz
&~~~~~~~~~&
(b) Grouped-layer ansatz with cyclic entangler
\end{tabular}
\caption{PQCs employed for numerical experiments.
Each layer consists of gates in the dashed line, and the total number of layers is written as $L$.
In sequential quantum optimization, parameterized single-qubit gates $R$ are updated in ascending order of the subscript.
}
\label{fig:alt}
\end{figure*}

\begin{figure*}[htb]
 \centering
 \begin{tabular}{ccc}
 \includegraphics[scale=0.28]{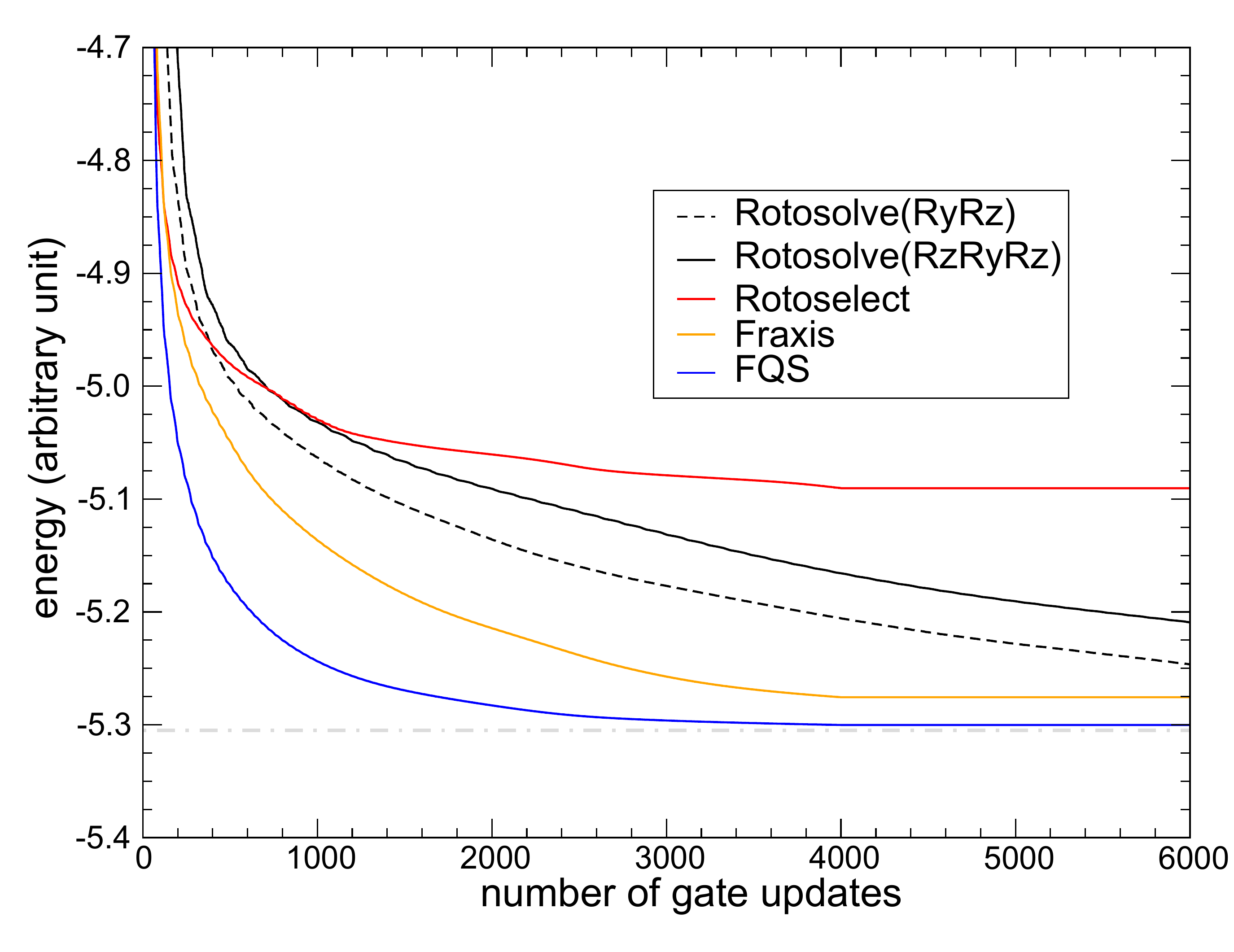}&&
  \includegraphics[scale=0.28]{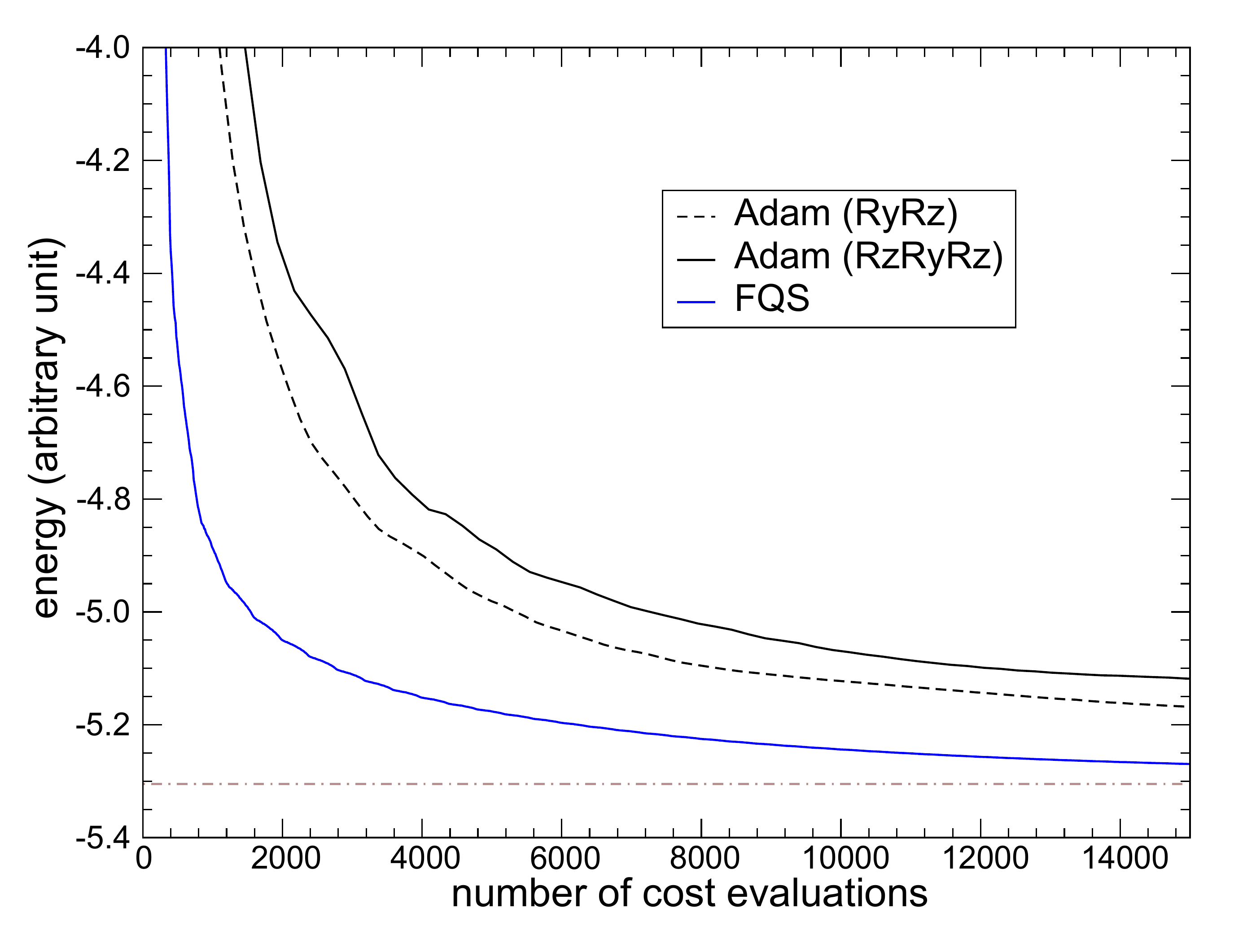}\\
  (a) Comparison with the other sequential optimizers
  &~~~&
  (b) Comparison with a gradient-based optimizer\\
 \end{tabular}
 \caption{Optimization trajectories of the VQE resulting energy for 5-qubit 1-dimensional mixed field Ising model on the 5-layer alternating layered ansatz, which is average over 20 independent runs.
 The horizontal dash-dotted line in gray color represents the exact ground energy.
 (left) Comparison with the other sequential optimizers, where the horizontal axis represents the number of gate updates.
  (right) Comparison with a gradient-based optimizer (Adam), where the horizontal axis represents the number of expectation evaluations. The learning rate parameter of 0.1 was employed for in the Adam optimizer.
 }
 \label{fig:Ising_trajectory}
\end{figure*}

\begin{figure*}[tb]
  \centering
  \begin{tabular}{ccc}
  \includegraphics[scale=0.28]{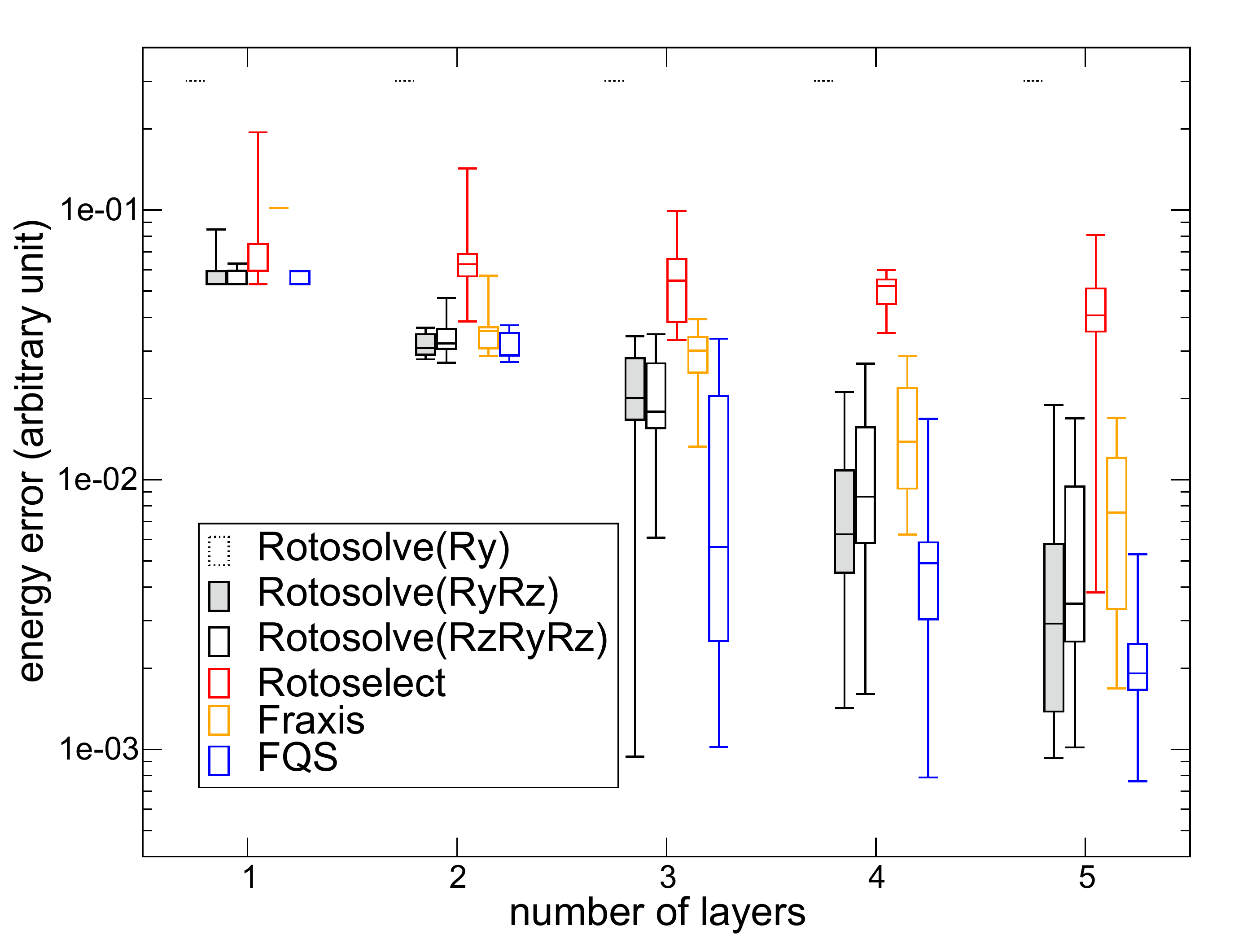}&&
  \includegraphics[scale=0.28]
  {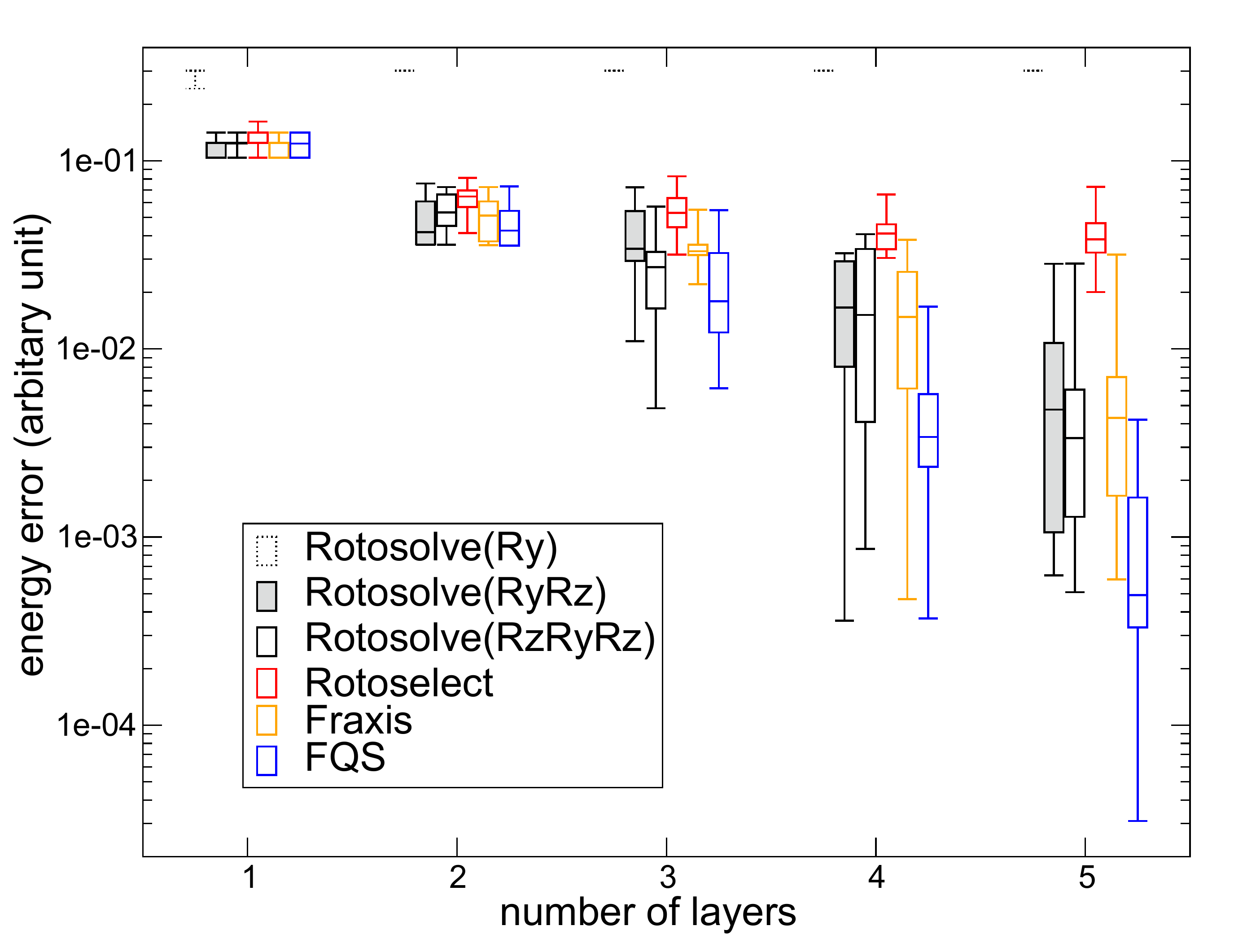}\\
  (a) Grouped-layer ansatz with cyclic entangler 
  &~~~&
  (b) Alternating layered ansatz\\
  \end{tabular}
  \caption{Boxplots of the resulting energy of VQE for 5-qubit 1-dimensional mixed field Ising model after 100 sweeps.
  The 20 independent VQEs were conducted from randomly-generated initial states.
  The vertical axis represents the difference between the obtained energy and the exact ground energy.
  }
 \label{fig:Ising_layers}
\end{figure*}
\noindent
We have seen that FQS generalizes known sequential optimizers such as NFT, Rotosolve/select, and Fraxis. 
Comparing with both other sequential quantum optimizers and a gradient-based optimizer, we show the advantages of FQS in VQE for the mixed field Ising model and fidelity maximization, where the former has a local Hamiltonian as the cost function while the latter is a global cost.
In addition, we demonstrate that our method can avoid the barren plateau (more precisely, the exponential concentration of spectrum of $S$) by numerically evaluating the second moment of spectral radius.
In the following, we provide numerical simulations based on the statevector simulator of Qiskit~\cite{Qiskit}.

\subsection{Mixed field Ising model}
To benchmark the performance of FQS, we carried out VQE optimizations for the 1-dimensional mixed field Ising model with five qubits whose Hamiltonian is 
\begin{align}\label{eq:ising}
H= J \sum_{i=1}^n Z^{(i)} Z^{(i+1)} 
+ h \sum_{i=1}^n (Y^{(i)}+Z^{(i)}),
\end{align}
where the superscript $i$ denotes the index of each site, and $J=1,h={1}/{\sqrt{2}}$. 
We employed the periodic boundary condition, that is, $Z^{(n+1)} = Z^{(1)}$.
We employed two types of ansatz as in Fig.~\ref{fig:alt}.
In the FQS optimizations, we prepared 20 initial parameter sets in the state-random manner, where the respective single-qubit gates were initialized based on the Haar random distribution.
Similarly, in the Fraxis optimizations, the rotational axis $\bm{n}$ in $R_{\bm{n}}(\psi)$ of each single-qubit gate was sampled from the uniform probability distribution on the Bloch sphere, while the rotational angles were fixed to $\pi$.
As for Rotoselect, the initial rotational axes were randomly selected from X, Y and Z axes, while the rotational angle were randomly initialized in both Rotosolve (NFT) and Rotoselect.

Figure~\ref{fig:Ising_trajectory}(a) shows averaged trajectories of independent 20 VQE simulations based on the 5-layer alternating layered ansatz.
Here we compare the convergence efficiency of FQS with those of the other sequential quantum optimizers with respect to the number of gate updates.
Although the number of circuit evaluations to update a single gate is not necessarily consistent among the optimizers, it may be a fair comparison because the practical wall times for a single gate update by these optimizers would be comparable if parallel circuit evaluations were allowed.

From Fig.~\ref{fig:Ising_trajectory}(a), it is obvious that FQS converged the most efficiently to the lowest energy value within 6000 updates.
Considering the gate expressibility, it may appear to be reasonable that FQS reached the better solution than Rotoselve and Fraxis.
We note, however, the efficiency of FQS is not necessarily trivial.
To verify it, we decomposed a general single-qubit gate, which we term an FQS gate, into equivalent three fixed-axis rotation gates $R_z(\phi)R_y(\vartheta)R_z(\lambda)$ (more precisely, we replaced $R$ in Fig.~\ref{fig:alt} with $RzRyRz$ gates) and sequentially optimized the three gates by Rotosolve.
In this case, the optimization turned out to be far slower not only than FQS but also than Fraxis. The ansatz with the three-gate decomposition was also slower than the $RyRz$ ansatz in optimization.
Actually, this deceleration of the $RzRyRz$ ansatz compared to $RyRz$ ansatz is consistent with the known dilemma; higher expressibility leads to lower trainability~\cite{holmes2022connecting}.
It is a contrast that FQS can maintain high optimization efficiency regardless of its high expressibility.
We can clearly attribute this efficiency of FQS to the incorporated parameter correlation within a single qubit gate, which has three degrees of freedom.
This insight is consistent with the fact that Fraxis also outperforms the sequential optimization for the $RyRz$ ansatz by Rotosolve, because Fraxis incorporates correlation of two degrees of freedom.

The comparison of performance between FQS and the gradient-based optimizers is presumably in great demand. 
However, it is not straightforward to compare them in a fair manner, because their apparent performances varies depending on the assumed hardware.
Figure~\ref{fig:Ising_trajectory}(b) shows a comparison of optimization efficiency between FQS and the gradient-based optimizer (Adam)~\cite{kingma2014adam}, where the horizontal axis represents the number of expectation evaluations, which is not consistent with that of Fig.~\ref{fig:Ising_trajectory}(a).
Although parallel computing is principally possible for the gradient-based optimizers, the required number of the cost evaluation for each optimization step is $O(D)$, where $D$ is the total number of gates in ansatz.
This is contrast to FQS that requires a constant number of the cost evaluation i.e., at most 10 circuits for a single gate update.
Since we did not find parallel computing practical for the gradient-based optimizations, we here employed the direct computational costs i.e., the number of expectation evaluations as alternative measure.
Figure~\ref{fig:Ising_trajectory}(b) clearly shows the FQS advantage over Adam optimizer.
The slower convergence of the Adam with the $RzRyRz$ gates can be attributed to the same reason for Rotosolve in Figure~\ref{fig:Ising_trajectory}(a).
Note that we employed the learning rate of 0.1 as a hyper-parameter for Adam, which appears to be rather larger than in its conventional usage in VQA, but provided modest results compared to 0.01 and 0.001 in benchmark simulations as shown Fig.~\ref{fig:adam_rate} in Appendix.

Figure~\ref{fig:Ising_layers} shows the optimized energy after 100 sweeps based on the ansaetze in Figs.~\ref{fig:alt}(a) and (b). 
In a single sweep, all gates were updated once in ascending order of the gate set index as labeled in Figs.~\ref{fig:alt}(a) and (b). Although the 100 sweeps may not necessarily be sufficient for rigorous convergence, the energy update by single sweep is too small to expect the drastic improvement by further iterations.
Figure~\ref{fig:Ising_layers} provides several insights as follows:
\begin{enumerate}
\item For all ansaetze employed, 
the optimizers did not show distinct difference except for Rotosolve with $Ry$ ansatz, if an extremely shallow circuit i.e., $L=1$.

\item FQS showed the best performance among all the sequential optimizers on all the ansaetze (alternating layered, cyclic, and ladder entanglers, see also Fig.~\ref{fig:Ising_layers_ladder} in  Appendix).
Especially, we confirmed a systematical improvement associated with the degrees of freedom of the target gate i.e., FQS optimizes 3 degrees, Fraxis optimizes 2 degrees and Rotosolve optimizes 1 degree.

\item The relative advantage of FQS over Rotosolve for the $RzRyRz$ ansatz depends on the circuit structure, where it is more distinct on the alternating layered ansatz rather than on the cyclic entangler ansatz.

\item The Rotosolve applications to a series of the fixed-axis gates ($RyRz$ and $RzRyRz$) were not better than FQS even though they have equivalent expressibility.
In some cases, on the contrary, a series of $RyRz$ gates showed better performance than Rotosolve with $RzRyRz$ gates that possess full expressibility for a single-qubit as well as an FQS gate.

\item The Rotoselect, which selects an optimal single-qubit gate within a part of $SU(2)$ unlike FQS, tended to be trapped in local minimum.

\end{enumerate}

Taking the higher expressibility and correlation among parameters of an FQS gate into account, the second insight apparently seems to be promising.
However, the third insight suggests that the higher circuit expressibility alone may not be sufficient for successful optimization, and on the contrary, in some cases it may disturb further optimization if correlation among parameters is not considered.
Note that barren plateaus do not account this hindrance because the FQS application that has equivalent expressibility to {\it RzRyRz} showed better performance.
Hence, we suppose that Rotosolve optimization for $RzRyRz$ gates is likely to be trapped at local minima and/or saddle points, while FQS may be more resilient due to its incorporating parameter correlation.

 \subsection{Fidelity maximization}\label{sec:fidelity_maximization}
\begin{figure}[!tb]
 \centering
      \includegraphics[scale=0.3]{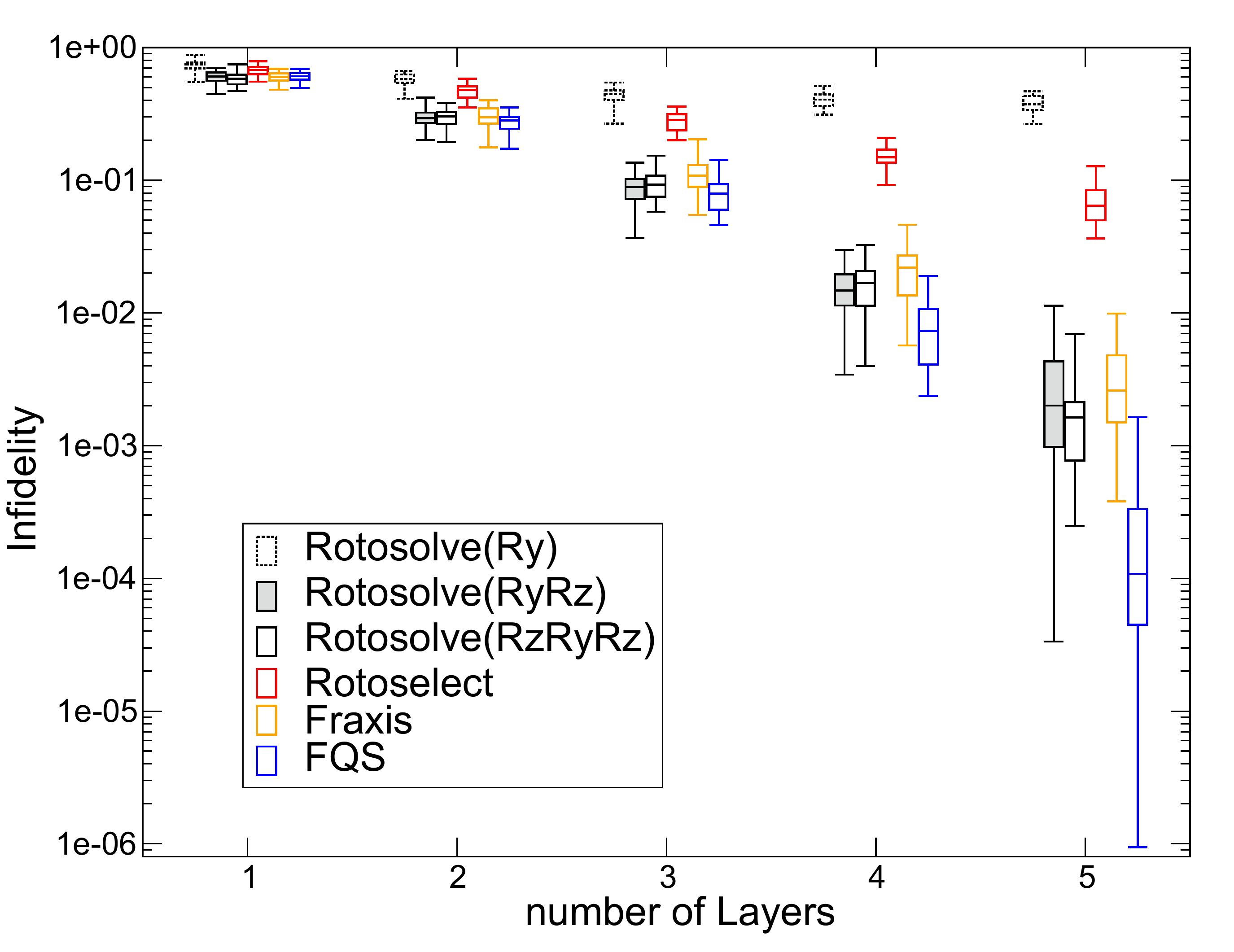}
 \caption{Boxplots of the VQE results for fidelity maximization as a function of the number of circuit layers. 
 The alternating layered ansatz in Fig.~\ref{fig:alt}(a) was employed.
 The box plot denotes quantiles consisting of independent 40 resulting energies with randomly-generated initial parameter set and target states taken from the Haar distribution.}
 \label{fig:fidelity}
 \end{figure}

 As another example to verify FQS performance, we conducted the fidelity maximization, where the infidelity with a reference state was regarded as the cost.
 The reference states were independently prepared with Haar random generator in Qiskit~\cite{Qiskit} in each optimization.
 Figure~\ref{fig:fidelity} shows the results after 100 sweeps of 40 independent runs where we employed the alternating layered ansatz as in Fig.~\ref{fig:alt}(a).
 We confirmed that FQS also showed better performance than the others, where the advantage was more distinct as the number of layers $L$ increased.
 This observation is consistent with the previous results on a local Hamiltonian.

\subsection{Spectral radius and noise-free barren plateau}
To verify our theorem on the spectrum of FQS matrix, we conducted numerical experiments by evaluating the second moment of the spectral radius of the centered FQS matrix Eq.~(\ref{eq:centered_S_matrix}), where an FQS matrix $S$ is evaluated with a randomly initialized quantum circuit.
Here, we employed the alternating layered ansatz in Fig.~\ref{fig:alt}(a) and two types of the cost function.
The first cost function is infidelity with a quantum state randomly generated with $n$-qubit Haar measure, which is a global cost function.
The second cost function is the expectation of the 1-local Hamiltonian defined as $H= Z\otimes I^{\otimes n-1}$.
\begin{figure*}[!tb]
 \centering
 \begin{tabular}{ccc}
 \includegraphics[scale=0.28]{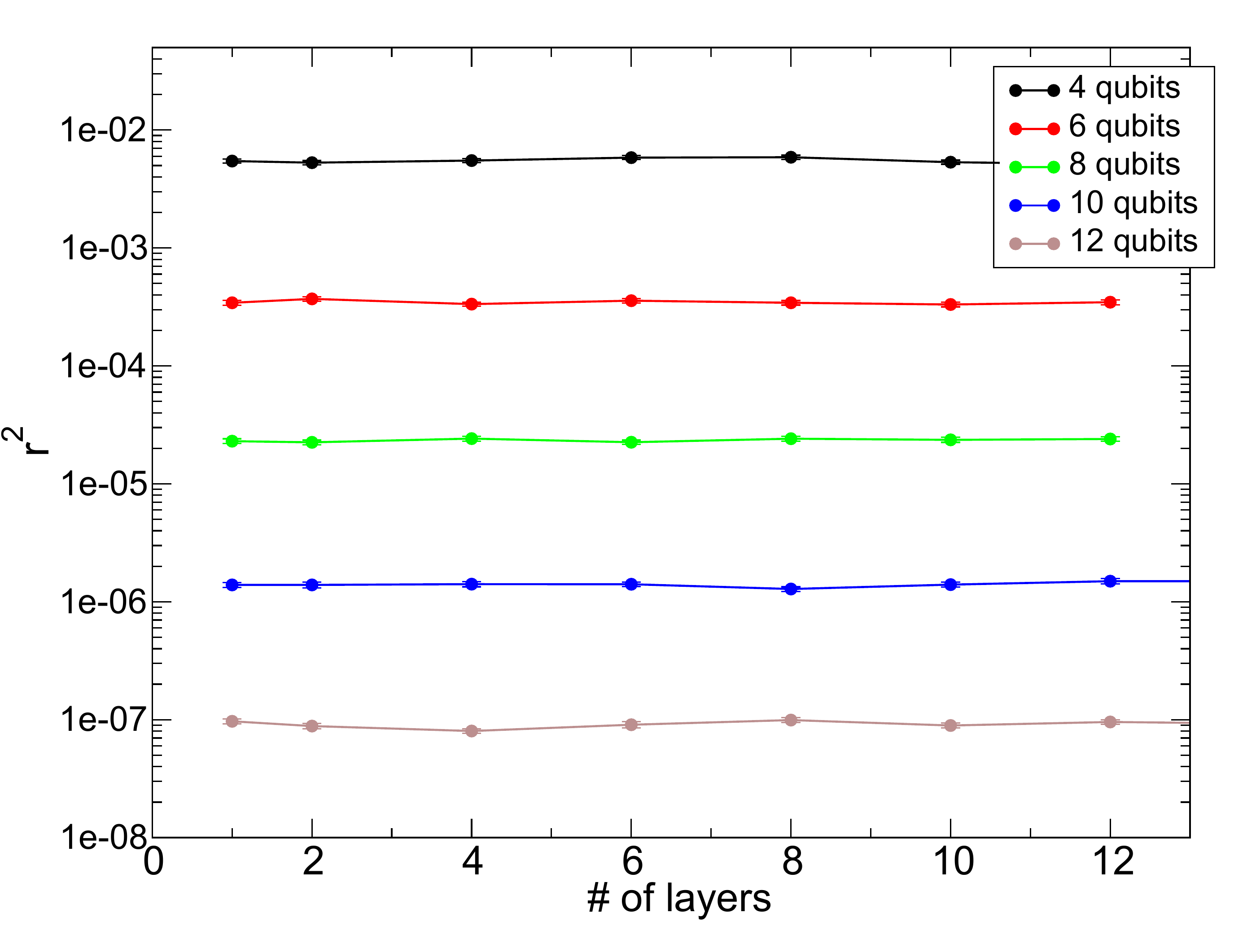}  
 &&
 \includegraphics[scale=0.28]{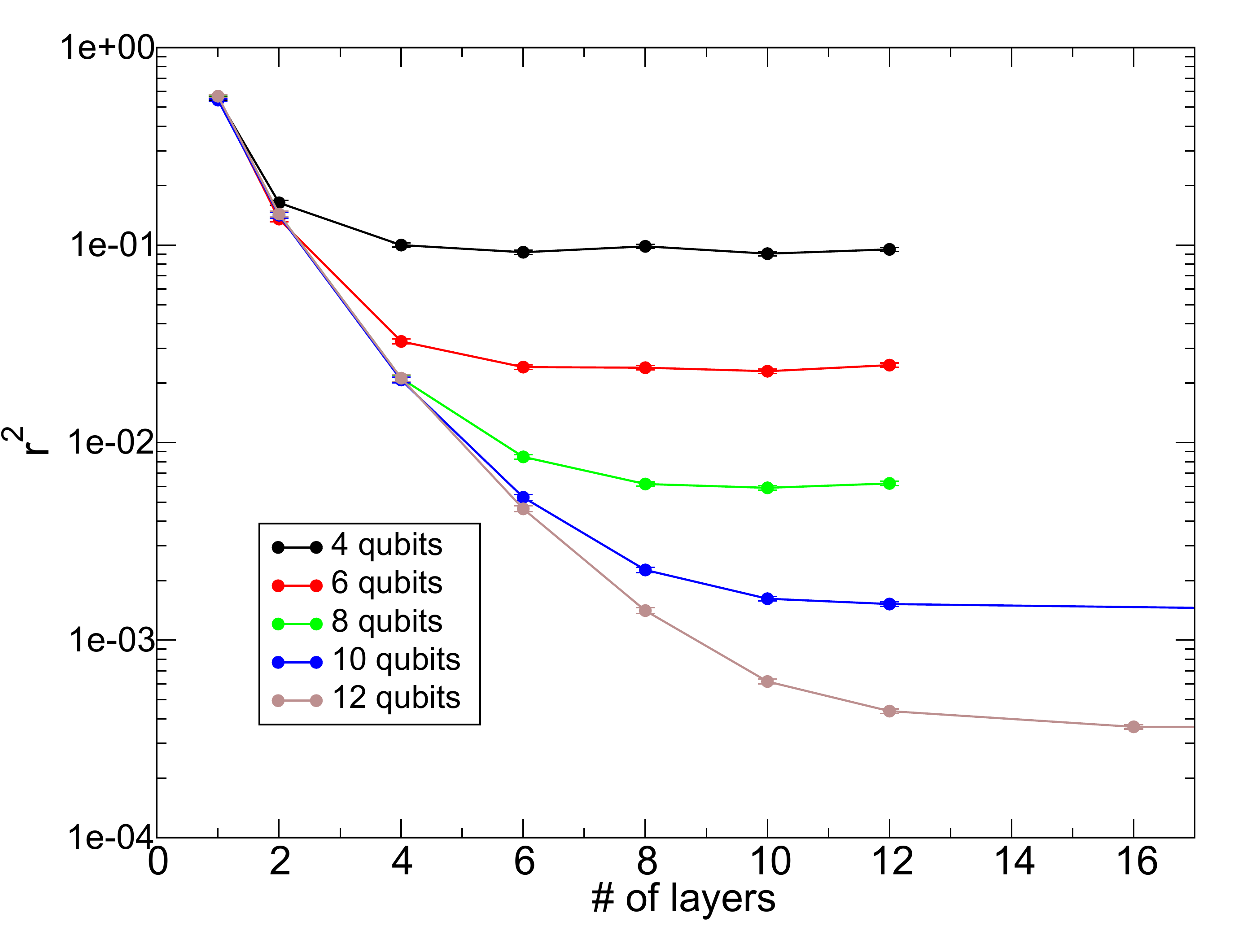} 
 \\
(a) Global cost  &&(b) Local cost \\
 \end{tabular}
 \caption{The second moment of spectral radius of the centered FQS matrix.
 The FQS matrix is evaluated for the single-qubit gate acting on the first qubit in the first layer of the alternating layered ansatz.
 The second moment is evaluated with 1000 samples based on the randomly initialized circuits with respect to the parameterized single-qubit gates.
 We employed the infidelity with the state generated using Haar measure on $n$-qubit unitary as the global cost.
 For the local cost function, we employed an expectation value of the Hamiltonian $H=Z\otimes I^{\otimes n-1}$.
 }
 \label{fig:spectrum_radius}
\end{figure*}

Figure~\ref{fig:spectrum_radius}(a) shows that the second moment is independent of the number of layers, but exponentially scaled according to the number of qubits.
Since the reference states generated with random unitary $U_{\rm Haar}$ from the unitary group with respect to the Haar measure are written as $\ket{\psi_{\rm ref}}=U_{\rm Haar}\ket{0}^{\otimes n}$, the fidelity can be regarded as the projection measurement on $\ket{0}^{\otimes n}$ with the input ansatz appended a randomly-generated unitary:
\begin{align}
 &|\braket{\psi_{\rm FQS}|\psi_{\rm ref}}|^2 \notag\\
 &~~= \bra{\psi_{\rm FQS}} U_{\rm Haar}\ket{0}\bra{0}^{\otimes n}U_{\rm Haar}^{\dagger} \ket{\psi_{\rm FQS}}.
\end{align}
As a result, the condition employed in Theorem~\ref{theorem1} holds, which is consistent with the present experiments.
In the case of the local cost function, the exponential decay of the second moment can be confirmed in the large limit of the number of layers in Fig.~\ref{fig:spectrum_radius}(b).
However, the second moment exhibited the transition from a constant value to exponentially small values as the number of layers increases.
Therefore, the sequential quantum optimizers are able to circumvent noise-free barren plateau using an alternating layered ansatz with limited number of layers for local cost functions.
We remark that although each block in the alternating layered ansatz used in the present experiments does not form unitary 2-design, the results seem to be consistent with the consequence of Theorem~\ref{theorem2}.
Therefore, the unitary 2-design in the block and even alternating layered ansatz may not be necessarily required to circumvent barren plateau.
Furthermore, we also observed the transition region of the second moment of the spectral radius with a different type of alternating layered ansatz (See Fig.~\ref{fig:spectral_radius_ladder} in Appendix).

\begin{figure}[t]
 \centering
 \begin{tabular}{c}
      \includegraphics[scale=0.30]{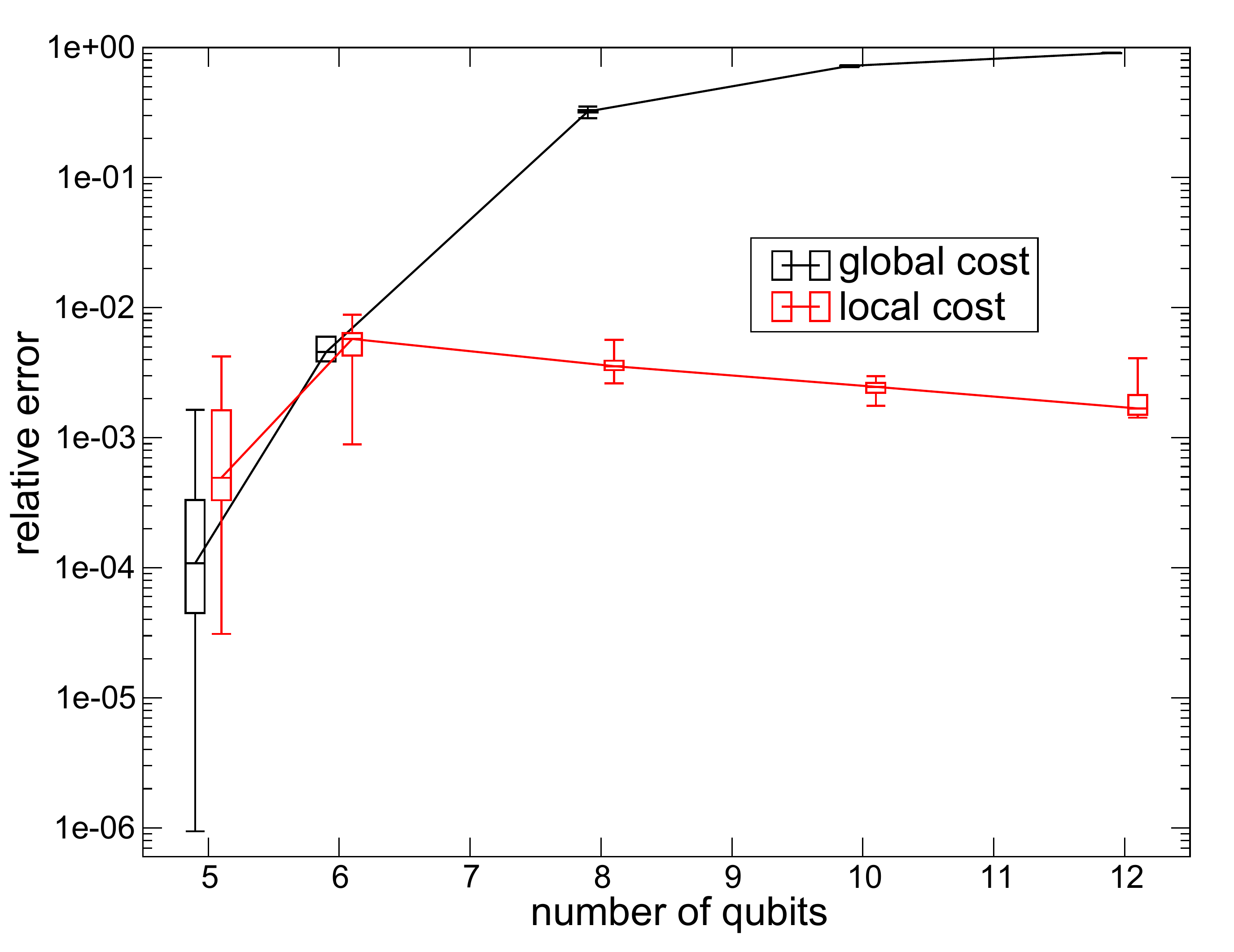}\\
 \end{tabular}
 \caption{Boxplots for the relative errors of the obtained cost measured from the exact solution after 100 sweeps as a function of the number of qubits. 
 As for the global cost, we executed the fidelity maximization as described in Sec.~\ref{sec:fidelity_maximization}, while the mixed field Ising Hamiltonian in Eq.~(\ref{eq:ising}) was employed as the local cost.
 For both VQEs, the 5-layer alternating layered ansatz in Fig.~\ref{fig:alt}(a) was employed.
 The box plots denote quantiles consisting of independent 20 optimizations.}
 \label{fig:check_avoid_BP}
\end{figure}

We also note that, at present, barren plateau has been analytically proven assuming the randomly initialized conditions.
Although it is the case at the beginning of the optimization, the randomness does not stand anymore during the optimization, and thus it is not trivial how the optimization proceeds after random initializations.
For further understanding, we carried on the additional optimizations with the 5-layer alternating layered ansatz varying the number of qubits.
Figure~\ref{fig:check_avoid_BP} shows the resulting relative errors after 100 sweeps of VQAs comparing the global and local cost functions.
Here, for the global cost we employed the fidelity with randomly-generated states as described in the previous section, while the local cost is the Hamiltonian of the mixed-field Ising model in Eq.~\eqref{eq:ising}. 
Note that the number of applications of FQS are not consistent in one sweep, because the total number of single-qubit gates increases according to the number of qubits.
The resulting errors for the global cost become exponentially larger as the number of qubits increases, which is in line with the spectral radius in the beginning of the optimizations as in Fig.~\ref{fig:spectrum_radius}(a).
In contrast, the results for local cost did not exhibit the exponential deterioration of trainability, which is also consistent with Fig.~\ref{fig:spectrum_radius}(b), where the relative error balanced after 6-qubit.
However, we expect that this error may increase again, as the number of qubits increases beyond a certain threshold, because a constant layer circuit does not have sufficient expressibility for the target states, even though the optimization proceeds to some extent in the beginning.
Thus, it is required for scalability to use a high expressible circuit e.g., alternating layered ansatz with high expressible blocks~\cite{nakaji2021expressibility} keeping the number of layers small.
Since our method utilizes the full expressiblity of single-qubit gates to optimize the circuit structure holding the number of layers, we believe that FQS is effective to this requirement.

\section{Conclusion} \label{sec:conclusion}
We proposed a new quantum algorithm for VQAs, called FQS, based on analytical optimization of circuit structure with respect to a single-qubit gate.
We have shown that the expectation of an observable on a quantum state prepared by a parameterized quantum circuit can be rewritten as the solvable quadratic form on parameters of a single-qubit gate in the cicuit, and our algorithm utilizes the matrix factorization based on this quadratic form.
The matrix factorization framework has also revealed the hierarchical relation of sequential quantum optimizers in the degrees of freedom of simultaneous optimization for single-qubit gate i.e., Rotosolve (NFT) $\leq$ Rotoselect $\leq$ Fraxis $\leq$ FQS.
Moreover, by introducing the spectral radius as a measure to evaluate the performance of sequential optimizers (including the existing methods), we rigorously proved the exponential scaling of spectrum concentration of the matrix associated with the quadratic form if the circuit is too deep, which is inherently equivalent to barren plateau in gradient optimizers.
On the other hand, we also proved the possibility to avoid the exponential concentration by supposing a local cost function and an alternating layered ansatz.

In the numerical experiments, we confirmed FQS achieved a good balance between trainability and expressibility by circuit structure optimization and incorporation of the intra-gate parameter correlation and outperformed a gradient-based optimizer as well as the other sequential optimizers.
We showed the efficacy of our proposed framework by extensive numerical experiments and confirmed the relations of sequential optimizers with barren plateaus.
Since FQS advances the circuit structure optimization from a heuristic to an analytical one, it may provide a solution to a critical problem of VQA, that is, circuit design. 
We hope the results are instrumental for promoting the use of sequential optimizers in VQAs.

\section{Acknowledgement}
K.W. would like to thank National Institute of Information and Communications Technology (NICT) for the support through Young Researchers Lab.
H.C.W. was supported by JSPS Grant Numbers
20K03885 and the MEXT Quantum Leap
Flagship Program Grant Number JPMXS0118067285
and JPMXS0120319794. 
We would like to thank Dr. Michihiko Sugawara, Dr. Yu-ya Ohnishi, Dr. Eriko Kaminishi, and Dr. Naoki Yamamoto for technical discussion.

\section{Code availability}
The FQS code used to create and the data in this article is available in Python using Qiskit version 0.36~\cite{Qiskit}, and can be found on GitHub at
\url{https://github.com/KQCC-Chemistry/SeQpt}.

\bibliographystyle{unsrt}
\bibliography{references}
\newpage
\onecolumngrid
\section{Appendix}

\renewcommand{\theequation}{A.\arabic{equation}}
\setcounter{equation}{0}
\subsection{Derivation of quadratic forms}\label{apdx:A}
Substituting the quaternion representation of single-qubit gates $R(\bm{q})=\bm{q}\cdot \vec{\varsigma}$ into the energy (\ref{eq:obj_fn}), we obtain
\begin{align}\label{apdx:quadratic_f}
    \langle H\rangle(\bm{q})&={\tr}\left[\rho'_{\rm in} R(\bm{q})^\dagger  H' R(\bm{q})\right]=\sum_{\mu,\nu=0}^3 {q}_{\mu}{q}_{\nu}{\rm tr}\left[\rho'_{\rm in}{\varsigma}^\dagger_{\mu} H'{\varsigma}_{\nu}\right]\notag\\
    &=\sum_{\mu,\nu=0}^3 A_{\mu\nu}{q}_{\mu}{q}_{\nu},~~~A_{\mu\nu}:={\rm tr}\left[\rho'_{\rm in}{\varsigma}^\dagger_{\mu} H'{\varsigma}_{\nu}\right]\notag\\
    &=\frac{1}{2}\left(\sum_{\mu,\nu=0}^3 A_{\mu\nu}{q}_{\mu}{q}_{\nu}+\sum_{\nu,\mu=0}^3 A_{\nu\mu}{q}_{\nu}{q}_{\mu}\right)=\frac{1}{2}\left(\sum_{\mu,\nu=0}^3 A_{\mu\nu}{q}_{\mu}{q}_{\nu}+\sum_{\mu,\nu=0}^3 A_{\nu\mu}{q}_{\mu}{q}_{\nu}\right)\notag\\
    &=\sum_{\mu,\nu=0}^3 \frac{A_{\mu\nu}+A_{\nu\mu}}{2}{q}_{\mu}{q}_{\nu}=\bm{q}^\top S\bm{q},
\end{align}
where the matrix $S:=(S_{\mu\nu})$ in the last line is defined by symmetrization of $A:=(A_{\mu\nu})$ as follows
\begin{align}
    S_{\mu\nu}&:=\frac{A_{\mu\nu}+A_{\nu\mu}}{2}=\frac12{\rm tr}\left[\rho'_{\rm in}\left({\varsigma}^\dagger_{\mu} H'{\varsigma}_{\nu}+{\varsigma}^\dagger_{\nu} H'{\varsigma}_{\mu}\right)\right].
\end{align}
Here, the $S$ is obviously a real symmetric matrix since ${\varsigma}^\dagger_{\mu} H'{\varsigma}_{\nu}+{\varsigma}^\dagger_{\nu} H'{\varsigma}_{\mu}$ is a Hermitian operator.

We write $\tilde{S}$ as the lower right $3\times 3$ part in $S$ matrix.
Then, the quadratic form for Fraxis (more precisely, $\pi$-Fraxis algorithm)~\cite{watanabe2021} is represented with $\tilde{S}$.
Actually, Fraxis algorithm can deal with the single-qubit gate $U_{\rm Fraxis}:=R_{\bm{n}}(\pi)$ and the corresponding quaternion is $q=(0,\bm{n})^\top$, which is simply located on the two-dimensional spherical surface.
Thus, from Eq.~(\ref{apdx:quadratic_f}) we directly obtain the quadratic form for Fraxis as
\begin{align}\label{apdx:Fraxis_quadratic}
    &{\tr}\left[\rho'_{\rm in} U_{\rm Fraxis}^\dagger  H' U_{\rm Fraxis}\right]=\bm{n}^\top\tilde{S}\bm{n}\notag\\
    &=\bm{n}^\top\begin{pmatrix}
    {\rm tr}(H'\sigma_1\rho'_{\rm in}\sigma_1)
    &\frac{1}{2}\left[{\rm tr}(H'\sigma_1\rho'_{\rm in}\sigma_2)
    +{\rm tr}(H'\sigma_2\rho'_{\rm in}\sigma_1)\right]
    &\frac{1}{2}\left[{\rm tr}(H'\sigma_1\rho'_{\rm in}\sigma_3)
    +{\rm tr}(H'\sigma_3\rho'_{\rm in}\sigma_1)\right]\\[5pt]
    \cdot&{\rm tr}(H'\sigma_2\rho'_{\rm in}\sigma_2)
    &\frac{1}{2}\left[{\rm tr}(H'\sigma_2\rho'_{\rm in}\sigma_3)
    +{\rm tr}(H'\sigma_3\rho'_{\rm in}\sigma_2)\right]\\[5pt]
    \cdot&\cdot&{\rm tr}(H'\sigma_3\rho'_{\rm in}\sigma_3)
    \end{pmatrix}\bm{n}.
\end{align}
As for the NFT~\cite{nakanishi2020} and Rotosolve/Rotoselect~\cite{ostaszewski2021}, they can optimize an axis-fixed single-qubit gate $U_{\rm NFT}:=R_{\bm{m}}(\psi)$, where $\bm{m}$ is a fixed rotational axis.
Substituting the quaternion $q=(\cos{\psi/2},\bm{m}\sin{\psi/2})^\top$ corresponding to this gate into the quadratic form of FQS, we obtain
\begin{align}\label{apdx:NFT_quadratic}
    {\tr}\left[\rho'_{\rm in} U_{\rm NFT}^\dagger  H' U_{\rm NFT}\right]&=S_{00}\cos^2\frac{\psi}{2}+\sum_{i=1,2,3}2S_{0i}m_i\sin\frac{\psi}{2}\cos\frac{\psi}{2}+\sum_{i,j=1,2,3}\sin^2\frac{\psi}{2}S_{ij}m_im_j\notag\\
    &=\begin{pmatrix}
    \cos\frac{\psi}2&\sin\frac{\psi}2
    \end{pmatrix}
    \begin{pmatrix}
    S_{00}&\vec{S}_0\cdot\boldsymbol{m}\\
    \vec{S}_0\cdot\boldsymbol{m}&\boldsymbol{m}^\top\tilde{S}\boldsymbol{m}\\
    \end{pmatrix}
    \begin{pmatrix}
    \cos\frac{\psi}2\\
    \sin\frac{\psi}2
    \end{pmatrix},
\end{align}
where $\vec{S_0}:=(S_{01},S_{02},S_{03})$.
Here, Rotoselect simply solve the minimization problems of the quadratic form for different axis $\bm{m}\in \Lambda$ and select the optimal axis for minimizing the energy, where $\Lambda$ is a predefined subset of rotational axis such as $\Lambda = \{(1,0,0), (0,1,0), (0,0,1)\}$.

\renewcommand{\theequation}{B.\arabic{equation}}
\setcounter{equation}{0}
\setcounter{thm}{0}
\subsection{Proof of main theorems}\label{apdx:B}
As derived in Appendix~\ref{apdx:A}, each sequential quantum optimizer has a $p\times p$ real-symmetric matrix. Here, $p-1$ is the degree of freedom of the target single-qubit gate ($-1$ means the constraint of normalization).
Note that FQS gate or general single-qubit gate $R(\bm{q})$ ($p=4$), Fraxis gate $R_{\bm{n}}(\pi)$ ($p=3$), and NFT gate with $\bm{m}$ axis $R_{\bm{m}}(\psi)$ ($p=2$) are written as the following unified form
\begin{align}
    R^{(p)}:=\sum_{\mu=0}^{p-1}q^{(p)}_{\mu} \varsigma^{(p)}_{\mu},
\end{align}
where $q^{(p)}_{\mu}$ is the $\mu$th element of unit vector in $\mathbb{R}^p$.
Here, $\varsigma^{(p)}_{\mu}$ is an extension of the Pauli matrices such as $(\vec{\varsigma})^{(3)}=(-iX,-iY,-iZ)$ for Fraxis gate and $(\vec{\varsigma})^{(2)}=(I,-i\bm{m}\cdot\vec{\sigma})$ for NFT gate.
Accordingly, each $p\times p$ real-symmetric matrix is also written as
\begin{align}
    S^{(p)}_{\mu\nu}:=\frac12{\rm tr}\left[\rho'_{\rm in}\left(\left({\varsigma^{(p)}_{\mu}}\right)^\dagger H'{\varsigma}^{(p)}_{\nu}+\left({\varsigma}^{(p)}_{\nu}\right)^\dagger H'{\varsigma}^{(p)}_{\mu}\right)\right].
\end{align}
Note that $S^{(3)}$ and $S^{(2)}$ correspond to the real-symmetric matrix in Eqs.~(\ref{apdx:Fraxis_quadratic}), (\ref{apdx:NFT_quadratic}), respectively.
In the below, we omit the superscript $p$ for simplicity. Moreover, it is convenient to write the elements of $S$ as
\begin{align}\label{apdx:S_simplified}
    S_{\mu\nu}&=\frac{1}{2}\sum_{(\alpha,\beta)=(\mu,\nu),(\nu,\mu)}{\rm tr}\left[HU_2\varsigma_{\alpha}U_1\rho_{\rm in}U_1^\dagger \varsigma^\dagger_{\beta}U_2^\dagger\right],~~~\mu,\nu=0,1,\cdots,p-1,\notag\\
    &=\frac{1}{2}\sum_{(\alpha,\beta)=(\mu,\nu),(\nu,\mu)}{\rm tr}\left[H^{(2)}\varsigma_{\alpha}\rho_{\rm in}^{(1)} \varsigma^\dagger_{\beta}\right],~~~H^{(2)}:=U_2^\dagger HU_2,~~~\rho_{\rm in}^{(1)}:=U_1\rho_{\rm in}U_1^\dagger.
\end{align}

\subsubsection{Proof of Theorem~\ref{theorem1}}
In the following, $d$ denotes the dimension of $n$-qubit system i.e., $d = 2^n$.
Here, we provide a proof of Theorem~\ref{theorem1}.
For convenience, we recall it.
\begin{thm}
    Suppose that the quantum circuits $U_1$ and $U_2$ are randomly and independently generated.
    If either $U_1$ or $U_2$ forms a unitary $t$-design with $t\geq 2$, the second moment of the spectral radius, $r$, for the centered matrix $S_{\rm c} = S-{\rm tr}[S]I_{p\times p}/p$ ($I_{p\times p}$ denotes the $p\times p$ identity) is upper bounded as follows.
    \begin{align}
        &\mathbb{E}_{U_1,U_2}\left[r\left(S_{\rm c}\right)^2\right]
        \leq\begin{dcases}
        \frac{p^2{\rm tr}\left[H^2\right]\Delta \rho_{\rm in,2}}{2(d^2-1)}+\frac{\Delta \rho_{\rm in,2}}{4p(d^2-1)}\sum_{\mu,\nu=0}^{p-1}\left(p(-1)^{1-\delta_{\mu\nu}}-2\right){\rm tr}\left[\mathbb{E}_{U_2}\left[H^{(2)}\varsigma_{\mu}\varsigma^\dagger_{\nu}H^{(2)}\varsigma_{\nu}\varsigma^\dagger_{\mu}+{\rm h.c.}\right]\right]\\
        \frac{p^2{\rm tr}\left[\rho_{\rm in}^2\right]\Delta H_2}{2(d^2-1)}+\frac{\Delta H_2}{4p(d^2-1)}\sum_{\mu,\nu=0}^{p-1}\left(p(-1)^{1-\delta_{\mu\nu}}-2\right){\rm tr}\left[\mathbb{E}_{U_1}\left[\rho_{\rm in}^{(1)} \varsigma^\dagger_{\mu}\varsigma_{\nu}\rho_{\rm in}^{(1)} \varsigma^\dagger_{\nu}\varsigma_{\mu}+{\rm h.c.}\right]\right]
        \end{dcases}
        ,
    \end{align}
    where the first case corresponds to $U_1$ being a $t$-design, and the second case corresponds to $U_2$ being a $t$-design.
    Here, $\mathbb{E}_{U_1,U_2}\left[\cdot\right]$ is defined as the expectation over the random quantum circuits $U_1$ and $U_2$, $\delta_{\mu\nu}$ denotes the Kronecker delta, {\rm h.c.} means the Hermite conjugate of the preceding term, and $\Delta \rho_{{\rm in},2},\Delta H_{2}$ are defined as
    \begin{align}
        \Delta \rho_{\rm in,2}:={\rm tr}\left[ \rho_{\rm in}^2\right]-\frac{1}{d},~~~\Delta H_2:={\rm tr}\left[ H^2\right]-\frac{{\rm tr}^2\left[ H\right]}{d}.
    \end{align}
\end{thm}
\begin{proof}
Since the centered matrix $S_{\rm c}$ is a real-symmetric matrix regardless of quantum circuits $U_1$ and $U_2$, the following inequality holds, which comes from the maximum is at most the square root of the sum of the squares.
\begin{align}
    r\left(S-\frac{{\rm tr}\left[S\right]}{p}I_{p\times p}\right)\leq \left\|S-\frac{{\rm tr}\left[S\right]}{p}I_{p\times p}\right\|_{\rm F},
\end{align}
where $\|\cdot\|_{\rm F}$ denotes the Frobenius norm.
Thus, the second moment of the spectral radius is evaluated as
\begin{align}\label{apdx:upperbound_r}
    \mathbb{E}_{U_1,U_2}\left[r\left(S-\frac{{\rm tr}\left[S\right]}{p}I_{p\times p}\right)^2\right]&\leq \mathbb{E}_{U_1,U_2}\left[\left\|S-\frac{{\rm tr}\left[S\right]}{p}I_{p\times p}\right\|_{\rm F}^2\right]\notag\\    &=\sum_{\mu\nu}\mathbb{E}_{U_1,U_2}\left[S^2_{\mu\nu}\right]-\frac{1}{p}\sum_{\mu\nu}\mathbb{E}_{U_1,U_2}\left[S_{\mu\mu}S_{\nu\nu}\right].
\end{align}

First, we evaluate the r.h.s of Eq.~(\ref{apdx:upperbound_r}) on the condition that the random quantum circuit $U_1$ forms a unitary $t$-design with $t\geq 2$.
Using lemma~\ref{lemma:U_1_haar}, which is proved in the next subsection, the following identity holds,
\begin{align}
    &\mathbb{E}_{U_1,U_2}\left[S_{\mu\nu}S_{\mu'\nu'}\right]\notag\\
    &=\frac{{\rm tr}^2\left[H\right]}{d^2-1}
    \Delta \rho_{\rm in,1}\delta_{\mu\nu}\delta_{\mu'\nu'}
    +\frac{1}{4(d^2-1)}\Delta \rho_{\rm in,2}\sum_{(\alpha,\beta)}\sum_{(\alpha',\beta')}\mathbb{E}_{U_2}\left[
    {\rm tr}\left[\varsigma^\dagger_{\beta}H^{(2)}\varsigma_{\alpha}\varsigma^\dagger_{\beta'}H^{(2)}\varsigma_{\alpha'}\right]\right].
\end{align}
where we defined
\begin{align}
    \Delta \rho_{\rm in,1}:=1-\frac{{\rm tr}\left[ \rho_{\rm in}^2\right]}{d}.
\end{align}
Hence, we can calculate the expectations in the final line of Eq.~(\ref{apdx:upperbound_r}) as follows.
\begin{align}
        \mathbb{E}_{U_1,U_2}\left[S_{\mu\nu}^2\right]
        &=\frac{{\rm tr}^2\left[H\right]}{d^2-1}
        \Delta \rho_{\rm in,1}\delta_{\mu\nu}\notag\\
        &~+\frac{1}{4(d^2-1)}\Delta \rho_{\rm in,2}\sum_{(\alpha,\beta)=(\mu,\nu),(\nu,\mu)}\sum_{(\alpha',\beta')=(\mu,\nu),(\nu,\mu)}\mathbb{E}_{U_2}\left[
        {\rm tr}\left[H^{(2)}\varsigma_{\alpha}\varsigma^\dagger_{\beta'}H^{(2)}\varsigma_{\alpha'}\varsigma^\dagger_{\beta}\right]\right]\notag\\
        &=\frac{{\rm tr}^2\left[H\right]}{d^2-1}
        \Delta \rho_{\rm in,1}\delta_{\mu\nu}+\frac{{\rm tr}\left[H^2\right]}{2(d^2-1)}\Delta \rho_{\rm in,2}\notag\\
        &~+\frac{1}{4(d^2-1)}\Delta \rho_{\rm in,2}
        {\rm tr}\left[\mathbb{E}_{U_2}\left[H^{(2)}\varsigma_{\mu}\varsigma^\dagger_{\nu}H^{(2)}\varsigma_{\mu}\varsigma^\dagger_{\nu}+{\rm h.c.}\right]\right],
\end{align}
\begin{align}
        \mathbb{E}_{U_1,U_2}\left[S_{\mu\mu}S_{\nu\nu}\right]
        &=\frac{{\rm tr}^2\left[H\right]}{d^2-1}
        \Delta \rho_{\rm in,1}\notag\\
        &~+\frac{1}{4(d^2-1)}\Delta \rho_{\rm in,2}\sum_{(\alpha,\beta)
        =(\mu,\mu),(\mu,\mu)}\sum_{(\alpha',\beta')=(\nu,\nu),(\nu,\nu)}\mathbb{E}_{U_2}\left[
        {\rm tr}\left[H^{(2)}\varsigma_{\alpha}\varsigma^\dagger_{\beta'}H^{(2)}\varsigma_{\alpha'}\varsigma^\dagger_{\beta}\right]\right]\notag\\
        &=\frac{{\rm tr}^2\left[H\right]}{d^2-1}
        \Delta \rho_{\rm in,1}+\frac{1}{d^2-1}\Delta \rho_{\rm in,2}\mathbb{E}_{U_2}\left[
        {\rm tr}\left[H^{(2)}\varsigma_{\mu}\varsigma^\dagger_{\nu}H^{(2)}\varsigma_{\nu}\varsigma^\dagger_{\mu}\right]\right]\notag\\
        &=\frac{{\rm tr}^2\left[H\right]}{d^2-1}
        \Delta \rho_{\rm in,1}+\frac{1}{2(d^2-1)}\Delta \rho_{\rm in,2}
        {\rm tr}\left[\mathbb{E}_{U_2}\left[H^{(2)}\varsigma_{\mu}\varsigma^\dagger_{\nu}H^{(2)}\varsigma_{\nu}\varsigma^\dagger_{\mu}+{\rm h.c.}\right]\right],
\end{align}
Accordingly, we can evaluate the second moment of the spectral radius as
\begin{align}
    &\mathbb{E}_{U_1,U_2}\left[S^2_{\mu\nu}\right]-\frac{1}{p}\mathbb{E}_{U_1,U_2}\left[S_{\mu\mu}S_{\nu\nu}\right]\notag\\
    &=\frac{{\rm tr}^2\left[H\right]}{p(d^2-1)}
    \Delta \rho_{\rm in,1}\left(p\delta_{\mu\nu}-1\right)
    +\frac{{\rm tr}\left[H^2\right]}{2(d^2-1)}\Delta \rho_{\rm in,2}\notag\\
    &~+\frac{\Delta \rho_{\rm in,2}}{4p(d^2-1)}\left(
    p{\rm tr}\left[\mathbb{E}_{U_2}\left[H^{(2)}\varsigma_{\mu}\varsigma^\dagger_{\nu}H^{(2)}\varsigma_{\mu}\varsigma^\dagger_{\nu}+{\rm h.c.}\right]\right]
    -
    2{\rm tr}\left[\mathbb{E}_{U_2}\left[H^{(2)}\varsigma_{\mu}\varsigma^\dagger_{\nu}H^{(2)}\varsigma_{\nu}\varsigma^\dagger_{\mu}+{\rm h.c.}\right]\right]\right)\notag\\
    &=\frac{{\rm tr}^2\left[H\right]}{p(d^2-1)}
    \Delta \rho_{\rm in,1}\left(p\delta_{\mu\nu}-1\right)
    +\frac{{\rm tr}\left[H^2\right]}{2(d^2-1)}\Delta \rho_{\rm in,2}\notag\\
    &~+\frac{\Delta \rho_{\rm in,2}}{4p(d^2-1)}
    \left(p(-1)^{1-\delta_{\mu\nu}}-2\right){\rm tr}\left[\mathbb{E}_{U_2}\left[H^{(2)}\varsigma_{\mu}\varsigma^\dagger_{\nu}H^{(2)}\varsigma_{\nu}\varsigma^\dagger_{\mu}+{\rm h.c.}\right]\right],
\end{align}
where we used $\varsigma_{\mu}\varsigma^\dagger_{\nu}=(-1)^{1-\delta_{\mu\nu}}\varsigma_{\nu}\varsigma^\dagger_{\mu}$ in the last equality, and
\begin{align}
    &\mathbb{E}_{U_1,U_2}\left[r\left(S-\frac{{\rm tr}\left[S\right]}{p}I_{p\times p}\right)^2\right]
    \leq\sum_{\mu\nu}\left(\mathbb{E}_{U_1,U_2}\left[S^2_{\mu\nu}\right]-\frac{1}{p}\mathbb{E}_{U_1,U_2}\left[S_{\mu\mu}S_{\nu\nu}\right]\right)\notag\\
    &=\frac{p^2{\rm tr}\left[H^2\right]\Delta \rho_{\rm in,2}}{2(d^2-1)}+\frac{{\rm tr}^2\left[H\right]\Delta \rho_{\rm in,1}}{p(d^2-1)}
    \sum_{\mu\nu}\left(p\delta_{\mu\nu}-1\right)\notag\\
    &~+\frac{\Delta \rho_{\rm in,2}}{4p(d^2-1)}
    \sum_{\mu\nu}\left(p(-1)^{1-\delta_{\mu\nu}}-2\right){\rm tr}\left[\mathbb{E}_{U_2}\left[H^{(2)}\varsigma_{\mu}\varsigma^\dagger_{\nu}H^{(2)}\varsigma_{\nu}\varsigma^\dagger_{\mu}+{\rm h.c.}\right]\right]\notag\\
    &=\frac{p^2{\rm tr}\left[H^2\right]\Delta \rho_{\rm in,2}}{2(d^2-1)}+\frac{\Delta \rho_{\rm in,2}}{4p(d^2-1)}
    \sum_{\mu\nu}\left(p(-1)^{1-\delta_{\mu\nu}}-2\right){\rm tr}\left[\mathbb{E}_{U_2}\left[H^{(2)}\varsigma_{\mu}\varsigma^\dagger_{\nu}H^{(2)}\varsigma_{\nu}\varsigma^\dagger_{\mu}+{\rm h.c.}\right]\right].
\end{align}

Next, we evaluate the r.h.s of Eq.~(\ref{apdx:upperbound_r}) on the condition that the random quantum circuit $U_2$ forms a unitary $t$-design with $t\geq 2$.
Using lemma~\ref{lemma:U_2_haar}, which is also proved in the next subsection, we calculate the expectations in the final line of Eq.~(\ref{apdx:upperbound_r}) as follows.
\begin{align}
        \mathbb{E}_{U_1,U_2}\left[S_{\mu\nu}^2\right]
        &=\frac{{\rm tr}^2\left[\rho_{\rm in}\right]}{d^2-1}\Delta H_1
        \delta_{\mu\nu}\notag\\
        &~+\frac{1}{4(d^2-1)}\Delta H_2\sum_{(\alpha,\beta)=(\mu,\nu),(\nu,\mu)}\sum_{(\alpha',\beta')=(\mu,\nu),(\nu,\mu)}\mathbb{E}_{U_1}\left[{\rm tr}\left[\rho_{\rm in}^{(1)} \varsigma^\dagger_{\beta}\varsigma_{\alpha'}\rho_{\rm in}^{(1)} \varsigma^\dagger_{\beta'}\varsigma_{\alpha}\right]\right]\notag\\
        &=\frac{1}{d^2-1}\Delta H_1
        \delta_{\mu\nu}+\frac{{\rm tr}\left[\rho_{\rm in}^2\right]}{2(d^2-1)}\Delta H_2\notag\\
        &~+\frac{1}{4(d^2-1)}\Delta H_2{\rm tr}\left[\mathbb{E}_{U_1}\left[\rho^{(1)}_{\rm in}\varsigma^\dagger_{\nu}\varsigma_{\mu}\rho^{(1)}_{\rm in}\varsigma_{\nu}^\dagger\varsigma_{\mu}+{\rm h.c.}\right]\right],
\end{align}
\begin{align}
        \mathbb{E}_{U_1,U_2}\left[S_{\mu\mu}S_{\nu\nu}\right]
        &=\frac{{\rm tr}^2\left[\rho_{\rm in}\right]}{d^2-1}\Delta H_1\notag\\
        &~+\frac{1}{4(d^2-1)}\Delta H_2\sum_{(\alpha,\beta)=(\mu,\mu),(\mu,\mu)}\sum_{(\alpha',\beta')=(\nu,\nu),(\nu,\nu)}\mathbb{E}_{U_1}\left[{\rm tr}\left[\rho_{\rm in}^{(1)} \varsigma^\dagger_{\beta}\varsigma_{\alpha'}\rho_{\rm in}^{(1)} \varsigma^\dagger_{\beta'}\varsigma_{\alpha}\right]\right]\notag\\
        &=\frac{1}{d^2-1}\Delta H_1
        +\frac{1}{d^2-1}\Delta H_2\mathbb{E}_{U_1}\left[{\rm tr}\left[\rho_{\rm in}^{(1)} \varsigma^\dagger_{\mu}\varsigma_{\nu}\rho_{\rm in}^{(1)} \varsigma^\dagger_{\nu}\varsigma_{\mu}\right]\right]\notag\\
        &=\frac{1}{d^2-1}\Delta H_1+\frac{1}{2(d^2-1)}\Delta H_2{\rm tr}\left[\mathbb{E}_{U_1}\left[\rho_{\rm in}^{(1)} \varsigma^\dagger_{\mu}\varsigma_{\nu}\rho_{\rm in}^{(1)} \varsigma^\dagger_{\nu}\varsigma_{\mu}+{\rm h.c.}\right]\right],
\end{align}
where we defined
\begin{align}
    \Delta H_1:={\rm tr}^2\left[ H\right]-\frac{{\rm tr}\left[ H^2\right]}{d}.
\end{align}
Accordingly, we can evaluate the second moment of the spectral radius as
\begin{align}
    &\mathbb{E}_{U_1,U_2}\left[S^2_{\mu\nu}\right]-\frac{1}{p}\mathbb{E}_{U_1,U_2}\left[S_{\mu\mu}S_{\nu\nu}\right]\notag\\
    &=\frac{1}{p(d^2-1)}\Delta H_1
    \left(p\delta_{\mu\nu}-1\right)+\frac{{\rm tr}\left[\rho_{\rm in}^2\right]}{2(d^2-1)}\Delta H_2\notag\\
    &~+\frac{\Delta H_2}{4p(d^2-1)}\left(p~{\rm tr}\left[\mathbb{E}_{U_1}\left[\rho^{(1)}_{\rm in}\varsigma^\dagger_{\nu}\varsigma_{\mu}\rho^{(1)}_{\rm in}\varsigma_{\nu}^\dagger\varsigma_{\mu}+{\rm h.c.}\right]\right]
    -2{\rm tr}\left[\mathbb{E}_{U_1}\left[\rho_{\rm in}^{(1)} \varsigma^\dagger_{\mu}\varsigma_{\nu}\rho_{\rm in}^{(1)} \varsigma^\dagger_{\nu}\varsigma_{\mu}+{\rm h.c.}\right]\right]\right)\notag\\
    &=\frac{1}{p(d^2-1)}\Delta H_1
    \left(p\delta_{\mu\nu}-1\right)+\frac{{\rm tr}\left[\rho_{\rm in}^2\right]}{2(d^2-1)}\Delta H_2\notag\\
    &~+\frac{\Delta H_2}{4p(d^2-1)}
    \left(p(-1)^{1-\delta_{\mu\nu}}-2\right){\rm tr}\left[\mathbb{E}_{U_1}\left[\rho_{\rm in}^{(1)} \varsigma^\dagger_{\mu}\varsigma_{\nu}\rho_{\rm in}^{(1)} \varsigma^\dagger_{\nu}\varsigma_{\mu}+{\rm h.c.}\right]\right],
\end{align}
where we used $\varsigma_{\nu}^\dagger\varsigma_{\mu}=(-1)^{1-\delta_{\mu\nu}}\varsigma_{\mu}^\dagger\varsigma_{\nu}$ in the last equality, and
\begin{align}
    &\mathbb{E}_{U_1,U_2}\left[r\left(S-\frac{{\rm tr}\left[S\right]}{p}I_{p\times p}\right)^2\right]
    \leq\sum_{\mu\nu}\left(\mathbb{E}_{U_1,U_2}\left[S^2_{\mu\nu}\right]-\frac{1}{p}\mathbb{E}_{U_1,U_2}\left[S_{\mu\mu}S_{\nu\nu}\right]\right)\notag\\
    &=\frac{p^2{\rm tr}\left[\rho_{\rm in}^2\right]\Delta H_2}{2(d^2-1)}+\frac{\Delta H_2}{4p(d^2-1)}
    \sum_{\mu\nu}\left(p(-1)^{1-\delta_{\mu\nu}}-2\right){\rm tr}\left[\mathbb{E}_{U_1}\left[\rho_{\rm in}^{(1)} \varsigma^\dagger_{\mu}\varsigma_{\nu}\rho_{\rm in}^{(1)} \varsigma^\dagger_{\nu}\varsigma_{\mu}+{\rm h.c.}\right]\right].
\end{align}
\end{proof}

\subsubsection{Proof of Theorem~\ref{theorem2}}

Here, we provide a proof of Theorem~\ref{theorem2}. For convenience, we recall it.
\begin{thm}
    Suppose that the whole quantum circuit $U$ is an $n$-qubit alternating layered ansatz with $m$-qubit blocks as described in Sec.~\ref{subsubsec:al_bp}.
    Here, we focus on a block $W$ in $l$th layer and a parameterized single-qubit gate $R$ in the block.
    We assume that the quantum circuits $W_{\rm A},W_{\rm B}\subset W$, which are located after and before the target gate respectively, and the other blocks form a local $2$-design independently.
    In addition, the Hamiltonian $H$ is assumed to be $m$-local such as
    \begin{align}\label{apdx:local_cost}
        H=c_0I^{\otimes n}+\sum_{i}c_i{h}_i,~c_0,c_i\in\mathbb{R},~(\exists i,c_i\neq 0),
        \end{align}
    where a tensor product of Pauli matrices $h_i$ acts non-trivially on at most $m$-qubit.
    (This is the same assumption in~\cite{Cerezo2021NatComm}.)
    Then, the second moment of the spectral radius is lower bounded as
    \begin{align}
        \mathbb{E}_{U_1,U_2}\left[r\left(S-\frac{{\rm tr}\left[S\right]}{p}I_{p\times p}\right)^2\right]&\geq \frac{(p+2)(p-1)2^{m(l+1)-1}}{p(2^{2m}-1)^2(2^{m}+1)^{L+l}} \sum_{i\in i_{\mathcal{L}}}\sum_{\substack{(k,k')\in k_{\mathcal{L}_{\rm B}}\\k'\geq k}} c_i^2\epsilon(\rho_{k,k'})\epsilon(h_i),
    \end{align}
    where $L$ denotes the total number of layers.
    Here, $i_{\mathcal{L}}$ is the set of $i$ indices whose associated operators $h_i$ act on qubits in the forward light-cone $\mathcal{L}$ of $W$, and $k_{\mathcal{L}_{\rm B}}$ is the set of $k$ indices whose associated subsystems $S_k$ are in the backward light-cone $\mathcal{L}_{\rm B}$ of $W$.
    The quantum state $\rho_{k,k'}$ is the reduced density matrix of the input state $\rho_{\rm in}$ on $S_k S_{k+1}\cdots S_{k'}$, and the function $\epsilon(M)$ for a matrix $M$ is defined as $\epsilon(M)=D_{\rm HS}(M,{\rm tr}(M)\mathbf{1}/d_M)$ where $D_{\rm HS}$ is the Hilbert-Schmidt distance and $d_M$ is the dimension of the matrix $M$.
\end{thm}
\begin{proof}
To establish the lower bound of the second moment, we begin with the following inequality, which comes from the maximum is at least the square root of the average of the sum of the squares.
\begin{align}
    r\left(S-\frac{{\rm tr}\left[S\right]}{p}I_{p\times p}\right)&\geq \frac{1}{\sqrt{p}}\left\|S-\frac{{\rm tr}\left[S\right]}{p}I_{p\times p}\right\|_{\rm F},
\end{align}
where $||\cdot||_{\rm F}$ denotes the Frobenius norm.
Since the inequality holds for any quantum circuits $U_1$ and $U_2$ in Fig.~\ref{fig:FQS_circuit},  
\begin{align}\label{apdx:lowerbound_r}
    \mathbb{E}_{U_1,U_2}\left[r\left(S-\frac{{\rm tr}\left[S\right]}{p}I_{p\times p}\right)^2\right]&\geq \frac{1}{p}\mathbb{E}_{U_1,U_2}\left[\left\|S-\frac{{\rm tr}\left[S\right]}{p}I_{p\times p}\right\|_{\rm F}^2\right]\notag\\
    &=\frac{1}{p}\sum_{\mu\nu}\left(\mathbb{E}_{U_1,U_2}\left[S^2_{\mu\nu}\right]-\frac{1}{p}\mathbb{E}_{U_1,U_2}\left[S_{\mu\mu}S_{\nu\nu}\right]\right).
\end{align}

We first evaluate the expectation of $S_{\mu\nu}S_{\mu'\nu'}$ over the block of interest.
From the setting, the block $W$ containing a single-qubit gate $R$ is decomposed as
\begin{align}
    W=W_{\rm A}(\mathbf{1}_{m-1}\otimes R)W_{\rm B},
\end{align}
where $\mathbf{1}_{m-1}$ is the identity on $m-1$ qubits system, and $W_{\rm A},W_{\rm B}\subset W$ are the quantum circuits after and before $R$, respectively.
As shown in Fig.~\ref{fig:FQS_circuit}, the quantum circuits $U_1$ and $U_2$ for an alternating layered ansatz can be written as
\begin{align}
    U_1=\left(\mathbf{1}_{\overline{w}}\otimes W_{\rm B}\right)V_1,~~~U_2=V_2\left(\mathbf{1}_{\overline{w}}\otimes W_{\rm A}\right),
\end{align}
where $\mathbf{1}_{\overline{w}}$ denotes the identity over the qubits on which the block $W$ acts trivially.
Here, $V_2$ contains the gates in the forward light-cone $\mathcal{L}$ of $W$, i.e., all gates with at least one input qubit causally connected to the qubits of $W$ as in Fig.~\ref{apdxfig:alansatz}, and $V_1$ contains other gates.
Accordingly, we can write the elements of $S$ matrix as
\begin{align}
    S_{\mu\nu}=\frac{1}{2}\sum_{(\alpha,\beta)=(\mu,\nu),(\nu,\mu)}{\rm tr}\left[HV_2\left(\mathbf{1}_{\overline{w}}\otimes W_{\rm A}(\mathbf{1}_{m-1}\otimes \varsigma_{\alpha})W_{\rm B}\right)V_1\rho_{\rm in} V_1^\dagger \left(\mathbf{1}_{\overline{w}}\otimes W_{\rm B}^\dagger(\mathbf{1}_{m-1}\otimes \varsigma^\dagger_{\beta})W_{\rm A}^\dagger\right)V_2^\dagger\right],
\end{align}
and we obtain
\begin{align}\label{apdx:ss_prod}
    &S_{\mu\nu}S_{\mu'\nu'}\notag\\
    &=\frac{1}{4}\sum_{\substack{(\alpha,\beta)\\(\alpha',\beta')}}
    {\rm tr}\left[HV_2\left(\mathbf{1}_{\overline{w}}\otimes W_{\rm A}(\mathbf{1}_{m-1}\otimes \varsigma_{\alpha})W_{\rm B}\right)V_1\rho_{\rm in} V_1^\dagger \left(\mathbf{1}_{\overline{w}}\otimes W_{\rm B}^\dagger(\mathbf{1}_{m-1}\otimes \varsigma^\dagger_{\beta})W_{\rm A}^\dagger\right)V_2^\dagger\right]\notag\\
    &~~~~~~~~~~~~~~~~~~\times {\rm tr}\left[HV_2\left(\mathbf{1}_{\overline{w}}\otimes W_{\rm A}(\mathbf{1}_{m-1}\otimes \varsigma_{\alpha'})W_{\rm B}\right)V_1\rho_{\rm in} V_1^\dagger \left(\mathbf{1}_{\overline{w}}\otimes W_{\rm B}^\dagger(\mathbf{1}_{m-1}\otimes \varsigma^\dagger_{\beta'})W_{\rm A}^\dagger\right)V_2^\dagger\right]\notag\\
    &=\frac{1}{4}\sum_{\substack{(\alpha,\beta)\\(\alpha',\beta')}} \sum_{\substack{\bm{i},\bm{j}\\\bm{i'},\bm{j'}}}{\rm tr}
    \left[W_{\rm A}(\mathbf{1}_{m-1}\otimes \varsigma_{\alpha})W_{\rm B}\rho^{(1)}_{{\rm in},\bm{ij}} W_{\rm B}^\dagger(\mathbf{1}_{m-1}\otimes \varsigma^\dagger_{\beta})W_{\rm A}^\dagger H^{(2)}_{\bm{ji}}\right]\notag\\
    &~~~~~~~~~~~~~~~~~~\times {\rm tr}\left[W_{\rm A}(\mathbf{1}_{m-1}\otimes \varsigma_{\alpha'})W_{\rm B}\rho^{(1)}_{{\rm in},\bm{i'j'}}W_{\rm B}^\dagger(\mathbf{1}_{m-1}\otimes \varsigma^\dagger_{\beta'})W_{\rm A}^\dagger  H^{(2)}_{\bm{j'i'}}\right].
\end{align}
Here, we defined
\begin{align}
    \rho_{{\rm in},\bm{ij}}^{(1)}:={\rm tr}_{\overline{w}}\left[\left(\ket{\bm{j}}\bra{\bm{i}}\otimes \mathbf{1}_\omega\right) V_1\rho_{\rm in}V_1^\dagger\right],
    ~~~
    H_{\bm{ji}}^{(2)}:={\rm tr}_{\overline{w}}\left[\left(\ket{\bm{i}}\bra{\bm{j}}\otimes \mathbf{1}_w \right)V_2^\dagger HV_2\right],
\end{align}
where ${\rm tr}_{\overline{\omega}}[\cdot]$ means the partial trace over the qubits which are not in $W$,
and $\{\ket{\bm{i}}\}$ denotes the computational basis on $(n-m)$-qubit system.

Since $W_{\rm A}$ forms a local $2$-design, we first compute the expectation of Eq.~(\ref{apdx:ss_prod}) over $W_{\rm A}$ as
\begin{align}\label{apdx:integrate_WA}
    &\mathbb{E}_{W_{\rm A}}\left[S_{\mu\nu}S_{\mu'\nu'}\right]\notag\\
    &=\frac{1}{4(2^{2m}-1)}\sum_{\substack{(\alpha,\beta)\\(\alpha',\beta')}} \sum_{\substack{\bm{i},\bm{j}\\\bm{i'},\bm{j'}}}
    \left({\rm tr}\left[H_{\bm{ji}}^{(2)}\right]{\rm tr}\left[H_{\bm{j'i'}}^{(2)}\right]-\frac{{\rm tr}\left[H_{\bm{ji}}^{(2)}H_{\bm{j'i'}}^{(2)}\right]}{2^m}\right)\notag\\
    &~~~~~~~~~~~~~~~~~~~~~~~\times{\rm tr}\left[(\mathbf{1}_{m-1}\otimes \varsigma_{\beta}^\dagger\varsigma_{\alpha})W_{\rm B}\rho^{(1)}_{{\rm in},\bm{ij}}W_{\rm B}^\dagger\right]{\rm tr}\left[(\mathbf{1}_{m-1}\otimes \varsigma_{\beta'}^\dagger\varsigma_{\alpha'})W_{\rm B}\rho^{(1)}_{{\rm in},\bm{i'j'}}W_{\rm B}^\dagger\right]\notag\\
    &~+\frac{1}{4(2^{2m}-1)}\sum_{\substack{(\alpha,\beta)\\(\alpha',\beta')}} \sum_{\substack{\bm{i},\bm{j}\\\bm{i'},\bm{j'}}}\left({\rm tr}\left[H_{\bm{ji}}^{(2)}H_{\bm{j'i'}}^{(2)}\right]-\frac{{\rm tr}\left[H_{\bm{ji}}^{(2)}\right]{\rm tr}\left[H_{\bm{j'i'}}^{(2)}\right]}{2^m}\right)\notag\\
    &~~~~~~~~~~~~~~~~~~~~~~~\times{\rm tr}\left[(\mathbf{1}_{m-1}\otimes \varsigma_{\beta'}^\dagger\varsigma_{\alpha})W_{\rm B}\rho^{(1)}_{{\rm in},\bm{ij}}W_{\rm B}^\dagger(\mathbf{1}_{m-1}\otimes \varsigma_{\beta}^\dagger\varsigma_{\alpha'})W_{\rm B}\rho^{(1)}_{{\rm in},\bm{i'j'}}W_{\rm B}^\dagger\right].
\end{align}
Here, we employed the following integration formula from the Weingarten calculus~\cite{Cerezo2021NatComm}
\begin{align}\label{apdx:weingarten_secondmoment}
\mathbb{E}_{W}~{\rm tr}\left[WAW^\dagger BWCW^\dagger D\right]&=\frac{1}{2^{2m}-1}\left({\rm tr}\left[A\right]{\rm tr}\left[C\right]{\rm tr}\left[BD\right]+{\rm tr}\left[AC\right]{\rm tr}\left[B\right]{\rm tr}\left[D\right]\right)\notag\\
&~~~-\frac{1}{2^m(2^{2m}-1)}\left({\rm tr}\left[AC\right]{\rm tr}\left[BD\right]+{\rm tr}\left[A\right]{\rm tr}\left[B\right]{\rm tr}\left[C\right]{\rm tr}\left[D\right]\right),
\end{align}
where $W$ is Haar-distributed on the unitary group of degree $2^m$, and $A,B,C$ and $D$ are arbitrary linear operators on an $m$-qubit system.
Noting that 
\begin{align}\label{apdx:vanish_WB}
    \sum_{(\alpha,\beta)=(\mu,\nu),(\nu,\mu)}{\rm tr}\left[(\mathbf{1}_{m-1}\otimes \varsigma_{\beta}^\dagger\varsigma_{\alpha})W_{\rm B}\rho^{(1)}_{{\rm in},\bm{ij}}W_{\rm B}^\dagger\right]=2\delta_{\mu\nu}{\rm tr}\left[\rho^{(1)}_{{\rm in},\bm{ij}}\right]
\end{align}
holds, then the first term of Eq.~(\ref{apdx:integrate_WA}) is independent of $W_{\rm B}$.
Since $W_{\rm B}$ also forms a local 2-design, a part of the second term of Eq.~(\ref{apdx:integrate_WA}) is evaluated as
\begin{align}\label{apdx:integrate_WB}
    &\sum_{\substack{(\alpha,\beta)\\(\alpha',\beta')}}\mathbb{E}_{W_{\rm B}}~{\rm tr}\left[(\mathbf{1}_{m-1}\otimes \varsigma_{\beta'}^\dagger\varsigma_{\alpha})W_{\rm B}\rho^{(1)}_{{\rm in},\bm{ij}}W_{\rm B}^\dagger(\mathbf{1}_{m-1}\otimes \varsigma_{\beta}^\dagger\varsigma_{\alpha'})W_{\rm B}\rho^{(1)}_{{\rm in},\bm{i'j'}}W_{\rm B}^\dagger\right]\notag\\
    &=\frac{1}{2^{2m}-1}\sum_{\substack{(\alpha,\beta)\\(\alpha',\beta')}}{\rm tr}\left[\mathbf{1}_{m-1}\otimes \varsigma_{\beta}^\dagger\varsigma_{\alpha'}\varsigma_{\beta'}^\dagger\varsigma_{\alpha}\right]\left({\rm tr}\left[\rho^{(1)}_{{\rm in},\bm{ij}}\right]{\rm tr}\left[\rho^{(1)}_{{\rm in},\bm{i'j'}}\right]-\frac{{\rm tr}\left[\rho^{(1)}_{{\rm in},\bm{ij}}\rho^{(1)}_{{\rm in},\bm{i'j'}}\right]}{2^m}\right)\notag\\
    &+\frac{1}{2^{2m}-1}\sum_{\substack{(\alpha,\beta)\\(\alpha',\beta')}}
    {\rm tr}\left[\mathbf{1}_{m-1}\otimes \varsigma_{\beta}^\dagger\varsigma_{\alpha'}\right]
    {\rm tr}\left[\mathbf{1}_{m-1}\otimes\varsigma_{\beta'}^\dagger\varsigma_{\alpha}\right]
    \left({\rm tr}\left[\rho^{(1)}_{{\rm in},\bm{ij}}\rho^{(1)}_{{\rm in},\bm{i'j'}}\right]-\frac{{\rm tr}\left[\rho^{(1)}_{{\rm in},\bm{ij}}\right]{\rm tr}\left[\rho^{(1)}_{{\rm in},\bm{i'j'}}\right]}{2^m}\right)\notag\\
    &=\frac{4\cdot 2^m}{2^{2m}-1}\delta_{\mu\nu}\delta_{\mu'\nu'}\left({\rm tr}\left[\rho^{(1)}_{{\rm in},\bm{ij}}\right]{\rm tr}\left[\rho^{(1)}_{{\rm in},\bm{i'j'}}\right]-\frac{{\rm tr}\left[\rho^{(1)}_{{\rm in},\bm{ij}}\rho^{(1)}_{{\rm in},\bm{i'j'}}\right]}{2^m}\right)\notag\\
    &~+\frac{8\cdot2^{2(m-1)}}{2^{2m}-1}\left(\delta_{\mu'\mu}\delta_{\nu'\nu}+\delta_{\mu\nu'}\delta_{\mu'\nu}\right)
    \left({\rm tr}\left[\rho^{(1)}_{{\rm in},\bm{ij}}\rho^{(1)}_{{\rm in},\bm{i'j'}}\right]-\frac{{\rm tr}\left[\rho^{(1)}_{{\rm in},\bm{ij}}\right]{\rm tr}\left[\rho^{(1)}_{{\rm in},\bm{i'j'}}\right]}{2^m}\right),
\end{align}
where we used the formula Eq.~(\ref{apdx:weingarten_secondmoment}) in the first equality.
In the second equality, we used the following relation, which can be derived from direct calculation, as
\begin{align}
    \sum_{(\alpha,\beta)=(\mu,\nu),(\nu,\mu)}\sum_{(\alpha',\beta')=(\mu',\nu'),(\nu',\mu')} {\rm tr}\left[\varsigma_{\beta}^\dagger\varsigma_{\alpha'}\right] {\rm tr}\left[\varsigma_{\beta'}^\dagger\varsigma_{\alpha}\right]=8(\delta_{\mu'\mu}\delta_{\nu'\nu}+\delta_{\mu\nu'}\delta_{\mu'\nu}).
\end{align}
Taking the expectation over $W_{\rm B}$ and substituting Eqs.~(\ref{apdx:vanish_WB}), (\ref{apdx:integrate_WB}) back into Eq.~(\ref{apdx:integrate_WA}), we obtain
\begin{align}\label{apdx:complete_int_W}
    &\mathbb{E}_{W_{\rm A},W_{\rm B}}\left[S_{\mu\nu}S_{\mu'\nu'}\right]\notag\\
    &=\frac{\delta_{\mu\nu}\delta_{\mu'\nu'}}{2^{2m}-1} \sum_{\substack{\bm{i},\bm{j}\\\bm{i'},\bm{j'}}}
    {\rm tr}\left[\rho^{(1)}_{{\rm in},\bm{ij}}\right]{\rm tr}\left[\rho^{(1)}_{{\rm in},\bm{i'j'}}\right]
    \left({\rm tr}\left[H_{\bm{ji}}^{(2)}\right]{\rm tr}\left[H_{\bm{j'i'}}^{(2)}\right]-\frac{{\rm tr}\left[H_{\bm{ji}}^{(2)}H_{\bm{j'i'}}^{(2)}\right]}{2^m}\right)\notag\\
    &~+\frac{2^m\delta_{\mu\nu}\delta_{\mu'\nu'}}{(2^{2m}-1)^2} \sum_{\substack{\bm{i},\bm{j}\\\bm{i'},\bm{j'}}}\left({\rm tr}\left[H_{\bm{ji}}^{(2)}H_{\bm{j'i'}}^{(2)}\right]-\frac{{\rm tr}\left[H_{\bm{ji}}^{(2)}\right]{\rm tr}\left[H_{\bm{j'i'}}^{(2)}\right]}{2^m}\right)\left({\rm tr}\left[\rho^{(1)}_{{\rm in},\bm{ij}}\right]{\rm tr}\left[\rho^{(1)}_{{\rm in},\bm{i'j'}}\right]-\frac{{\rm tr}\left[\rho^{(1)}_{{\rm in},\bm{ij}}\rho^{(1)}_{{\rm in},\bm{i'j'}}\right]}{2^m}\right)\notag\\
    &~+\frac{2\cdot 2^{2m}}{4(2^{2m}-1)^2} \left(\delta_{\mu'\mu}\delta_{\nu'\nu}+\delta_{\mu\nu'}\delta_{\mu'\nu}\right)\notag\\
    &~~~~~\times\sum_{\substack{\bm{i},\bm{j}\\\bm{i'},\bm{j'}}}\left({\rm tr}\left[H_{\bm{ji}}^{(2)}H_{\bm{j'i'}}^{(2)}\right]-\frac{{\rm tr}\left[H_{\bm{ji}}^{(2)}\right]{\rm tr}\left[H_{\bm{j'i'}}^{(2)}\right]}{2^m}\right)
    \left({\rm tr}\left[\rho^{(1)}_{{\rm in},\bm{ij}}\rho^{(1)}_{{\rm in},\bm{i'j'}}\right]-\frac{{\rm tr}\left[\rho^{(1)}_{{\rm in},\bm{ij}}\right]{\rm tr}\left[\rho^{(1)}_{{\rm in},\bm{i'j'}}\right]}{2^m}\right)\notag\\
    &=\delta_{\mu\nu}\delta_{\mu'\nu'}T^{(1,2)}+\frac{2\cdot 2^{2m}}{4(2^{2m}-1)^2} \left(\delta_{\mu'\mu}\delta_{\nu'\nu}+\delta_{\mu\nu'}\delta_{\mu'\nu}\right)\sum_{\substack{\bm{i},\bm{j}\\\bm{i'},\bm{j'}}}\Delta H_{\bm{ij}}^{\bm{i'j'}}\Delta \rho_{\bm{ij}}^{\bm{i'j'}}.
\end{align}
Here, we defined
\begin{align}
    \Delta H_{\bm{ij}}^{\bm{i'j'}}:={\rm tr}\left[H_{\bm{ji}}^{(2)}H_{\bm{j'i'}}^{(2)}\right]-\frac{{\rm tr}\left[H_{\bm{ji}}^{(2)}\right]{\rm tr}\left[H_{\bm{j'i'}}^{(2)}\right]}{2^m},
\end{align}
\begin{align}
    \Delta \rho_{\bm{ij}}^{\bm{i'j'}}:={\rm tr}\left[\rho^{(1)}_{{\rm in},\bm{ij}}\rho^{(1)}_{{\rm in},\bm{i'j'}}\right]-\frac{{\rm tr}\left[\rho^{(1)}_{{\rm in},\bm{ij}}\right]{\rm tr}\left[\rho^{(1)}_{{\rm in},\bm{i'j'}}\right]}{2^m},
\end{align}
\begin{align}
    &T^{(1,2)}:=\frac{1}{2^{2m}-1} \sum_{\substack{\bm{i},\bm{j}\\\bm{i'},\bm{j'}}}
    {\rm tr}\left[\rho^{(1)}_{{\rm in},\bm{ij}}\right]{\rm tr}\left[\rho^{(1)}_{{\rm in},\bm{i'j'}}\right]
    \left({\rm tr}\left[H_{\bm{ji}}^{(2)}\right]{\rm tr}\left[H_{\bm{j'i'}}^{(2)}\right]-\frac{{\rm tr}\left[H_{\bm{ji}}^{(2)}H_{\bm{j'i'}}^{(2)}\right]}{2^m}\right)\notag\\
    &~+\frac{2^m}{(2^{2m}-1)^2} \sum_{\substack{\bm{i},\bm{j}\\\bm{i'},\bm{j'}}}
    \Delta H_{\bm{ij}}^{\bm{i'j'}}
    \left({\rm tr}\left[\rho^{(1)}_{{\rm in},\bm{ij}}\right]{\rm tr}\left[\rho^{(1)}_{{\rm in},\bm{i'j'}}\right]-\frac{{\rm tr}\left[\rho^{(1)}_{{\rm in},\bm{ij}}\rho^{(1)}_{{\rm in},\bm{i'j'}}\right]}{2^m}\right).
\end{align}

Before proceeding to evaluate the expectation over $V_1$ and $V_2$, we calculate the summation in Eq.~(\ref{apdx:lowerbound_r}) as follows.
\begin{align}
    \sum_{\mu\nu}\mathbb{E}_{W_{\rm A},W_{\rm B}}\left[S^2_{\mu\nu}\right]&=pT^{(1,2)}+\frac{2\cdot 2^{2m}}{4(2^{2m}-1)^2} \left(p^2+p\right)\sum_{\substack{\bm{i},\bm{j}\\\bm{i'},\bm{j'}}}\Delta H_{\bm{ij}}^{\bm{i'j'}}\Delta \rho_{\bm{ij}}^{\bm{i'j'}},
\end{align}
\begin{align}
    \sum_{\mu\nu}\mathbb{E}_{W_{\rm A},W_{\rm B}}\left[S_{\mu\mu}S_{\nu\nu}\right]&=p^2T^{(1,2)}+\frac{2^{2m}}{(2^{2m}-1)^2}p \sum_{\substack{\bm{i},\bm{j}\\\bm{i'},\bm{j'}}}\Delta H_{\bm{ij}}^{\bm{i'j'}}\Delta \rho_{\bm{ij}}^{\bm{i'j'}}.
\end{align}
Accordingly,
\begin{align}\label{apdx:lower_bound_pre}
    \sum_{\mu\nu}\left(\mathbb{E}_{U_1,U_2}\left[S^2_{\mu\nu}\right]-\frac{1}{p}\mathbb{E}_{U_1,U_2}\left[S_{\mu\mu}S_{\nu\nu}\right]\right)
    &=\mathbb{E}_{V_1,V_2}\left[\sum_{\mu\nu}\mathbb{E}_{W_{\rm A},W_{\rm B}}\left[S^2_{\mu\nu}\right]-\frac{1}{p}\sum_{\mu\nu}\mathbb{E}_{W_{\rm A},W_{\rm B}}\left[S_{\mu\mu}S_{\nu\nu}\right]\right]\notag\\
    &=\frac{2\cdot 2^{2m}(p+2)(p-1)}{4(2^{2m}-1)^2}\sum_{\substack{\bm{i},\bm{j}\\\bm{i'},\bm{j'}}}
    \mathbb{E}_{V_1}\left[\Delta \rho_{\bm{ij}}^{\bm{i'j'}}\right]
    \mathbb{E}_{V_2}\left[\Delta H_{\bm{ij}}^{\bm{i'j'}}\right].
\end{align}

Finally, we evaluate the expectations over $V_1,V_2$ in Eq.~(\ref{apdx:lower_bound_pre}), which can be calculated basically with a series of integration for the $m$-qubit blocks in $V_1,V_2$.
In the same assumption of ours, the previous study~\cite{Cerezo2021NatComm} has showed the following inequality holds:
\begin{align}
    \sum_{\substack{\bm{i},\bm{j}\\\bm{i'},\bm{j'}}}
    \mathbb{E}_{V_1}\left[\Delta \rho_{\bm{ij}}^{\bm{i'j'}}\right]
    \mathbb{E}_{V_2}\left[\Delta H_{\bm{ij}}^{\bm{i'j'}}\right]\geq \frac{2^{m(l-1)}}{(2^{m}+1)^{L+l}}\sum_{i\in i_{\mathcal{L}}}\sum_{\substack{(k,k')\in k_{\mathcal{L}_{\rm B}}\\k'\geq k}} c_i^2\epsilon(\rho_{k,k'})\epsilon(h_i),
\end{align}
where we recall that $L$ is the total number of layers, and the block $W$ of interest is in the $l$th layer.
Combining this inequality with Eq.~(\ref{apdx:lower_bound_pre}), we establish the lower bound for the second moment of spectral radius as
\begin{align}
    \mathbb{E}_{U_1,U_2}\left[r\left(S-\frac{{\rm tr}\left[S\right]}{p}I_{p\times p}\right)^2\right]&\geq\frac{1}{p}\sum_{\mu\nu}\left(\mathbb{E}_{U_1,U_2}\left[S^2_{\mu\nu}\right]-\frac{1}{p}\mathbb{E}_{U_1,U_2}\left[S_{\mu\mu}S_{\nu\nu}\right]\right)\notag\\
    &=\frac{1}{p}\frac{2\cdot 2^{2m}(p+2)(p-1)}{4(2^{2m}-1)^2}\sum_{\substack{\bm{i},\bm{j}\\\bm{i'},\bm{j'}}}
    \mathbb{E}_{V_1}\left[\Delta \rho_{\bm{ij}}^{\bm{i'j'}}\right]
    \mathbb{E}_{V_2}\left[\Delta H_{\bm{ij}}^{\bm{i'j'}}\right]\notag\\
    &\geq \frac{(p+2)(p-1)2^{m(l+1)-1}}{p(2^{2m}-1)^2(2^{m}+1)^{L+l}} \sum_{i\in i_{\mathcal{L}}}\sum_{\substack{(k,k')\in k_{\mathcal{L}_{\rm B}}\\k'\geq k}} c_i^2\epsilon(\rho_{k,k'})\epsilon(h_i).
\end{align}
\end{proof}

\renewcommand{\theequation}{C.\arabic{equation}}
\setcounter{equation}{0}
\subsection{Proof of some useful lemmas}
In this subsection, we provide some lemmas for the proof of Theorem~\ref{theorem1}.
\begin{lemma}\label{lemma:U_1_haar}
    Suppose that the random quantum circuits $U_1$ and $U_2$ are independent.
    If $U_1$ forms a unitary $t$-design with $t\geq 2$, the expectation of the product of two matrix elements for $S$ can be evaluated as
    \begin{align}\label{apdx:lemma1_res}
      \mathbb{E}_{U_1,U_2}\left[S_{\mu\nu}S_{\mu'\nu'}\right]
        &=\frac{{\rm tr}^2\left[H\right]}{d^2-1}
        \Delta \rho_{\rm in,1}\delta_{\mu\nu}\delta_{\mu'\nu'}
        +\frac{1}{4(d^2-1)}\Delta \rho_{\rm in,2}\sum_{(\alpha,\beta)}\sum_{(\alpha',\beta')}\mathbb{E}_{U_2}\left[
        {\rm tr}\left[\varsigma^\dagger_{\beta}H^{(2)}\varsigma_{\alpha}\varsigma^\dagger_{\beta'}H^{(2)}\varsigma_{\alpha'}\right]\right].
    \end{align}
    Here, the summation runs over the set of $\mu,\nu$ as follows
    \begin{align}
        \sum_{(\alpha,\beta)}\sum_{(\alpha',\beta')}=\sum_{(\alpha,\beta)=(\mu,\nu),(\nu,\mu)}\sum_{(\alpha',\beta')=(\mu',\nu'),(\nu',\mu')}.
    \end{align}
\end{lemma}

\begin{proof}
Substituting Eq.~(\ref{apdx:S_simplified}) into the l.h.s of Eq.~(\ref{apdx:lemma1_res}) and omitting the expectation over $U_2$, we obtain
\begin{align}\label{apdx:U_1_exp_SS}
    &\mathbb{E}_{U_1}\left[S_{\mu\nu}S_{\mu'\nu'}\right]\notag\\
    &=\frac{1}{4}\sum_{(\alpha,\beta)=(\mu,\nu),(\nu,\mu)}\sum_{(\alpha',\beta')=(\mu',\nu'),(\nu',\mu')}\mathbb{E}_{U_1}~{\rm tr}\left[H^{(2)}\varsigma_{\alpha}U_1\rho_{\rm in}U_1^\dagger \varsigma^\dagger_{\beta}\right]{\rm tr}\left[H^{(2)}\varsigma_{\alpha'}U_1\rho_{\rm in}U_1^\dagger \varsigma^\dagger_{\beta'}\right]\notag\\
    &=\frac{1}{4}\sum_{(\alpha,\beta)}\sum_{(\alpha',\beta')}\sum_{\bm{i},\bm{j}}~\mathbb{E}_{U_1}~{\rm tr}\left[U_1\rho_{\rm in}U_1^\dagger \varsigma^\dagger_{\beta}H^{(2)}\varsigma_{\alpha}\ket{\bm{i}}\bra{\bm{j}}U_1\rho_{\rm in}U_1^\dagger \varsigma^\dagger_{\beta'}H^{(2)}\varsigma_{\alpha'}\ket{\bm{j}}\bra{\bm{i}}\right],
\end{align}
where $\{\ket{\bm{i}}\}$ denotes the computational basis on $n$-qubit system.
If $U_1$ forms a unitary $t$-design with $t\geq 2$, we can evaluate
\begin{align}\label{apdx:transform_wg_U1}
    &\sum_{\bm{i},\bm{j}}~\mathbb{E}_{U_1}~{\rm tr}\left[U_1\rho_{\rm in}U_1^\dagger \varsigma^\dagger_{\beta}H^{(2)}\varsigma_{\alpha}\ket{\bm{i}}\bra{\bm{j}}U_1\rho_{\rm in}U_1^\dagger \varsigma^\dagger_{\beta'}H^{(2)}\varsigma_{\alpha'}\ket{\bm{j}}\bra{\bm{i}}\right]\notag\\
    &=\frac{1}{d^2-1}\left({\rm tr}\left[\varsigma^\dagger_{\beta}H^{(2)}\varsigma_{\alpha}\right]{\rm tr}\left[\varsigma^\dagger_{\beta'}H^{(2)}\varsigma_{\alpha'}\right]+{\rm tr}\left[\rho_{\rm in}^2\right]{\rm tr}\left[\varsigma^\dagger_{\beta}H^{(2)}\varsigma_{\alpha}\varsigma^\dagger_{\beta'}H^{(2)}\varsigma_{\alpha'}\right]\right)\notag\\
    &~-\frac{1}{d(d^2-1)}\left({\rm tr}\left[\rho_{\rm in}^2\right]{\rm tr}\left[\varsigma^\dagger_{\beta}H^{(2)}\varsigma_{\alpha}\right]{\rm tr}\left[\varsigma^\dagger_{\beta'}H^{(2)}\varsigma_{\alpha'}\right]
    +{\rm tr}\left[\varsigma^\dagger_{\beta}H^{(2)}\varsigma_{\alpha}\varsigma^\dagger_{\beta'}H^{(2)}\varsigma_{\alpha'}\right]\right),
\end{align}
where we used the formula in Eq.~(\ref{apdx:weingarten_secondmoment}).
Substituting Eq.~(\ref{apdx:transform_wg_U1}) back into Eq.~(\ref{apdx:U_1_exp_SS}), we obtain
\begin{align}\label{apdx:U_1_haar_pre}
    &\mathbb{E}_{U_1}\left[S_{\mu\nu}S_{\mu'\nu'}\right]\notag\\
    &=\frac{1}{4}\sum_{(\alpha,\beta)}\sum_{(\alpha',\beta')}~\frac{1}{d^2-1}\left({\rm tr}\left[\varsigma^\dagger_{\beta}H^{(2)}\varsigma_{\alpha}\right]{\rm tr}\left[\varsigma^\dagger_{\beta'}H^{(2)}\varsigma_{\alpha'}\right]\left(1-\frac{{\rm tr}\left[\rho_{\rm in}^2\right]}{d}\right)\right)\notag\\
    &~+\frac{1}{4}\sum_{(\alpha,\beta)}\sum_{(\alpha',\beta')}~\frac{1}{d^2-1}\left(
    {\rm tr}\left[\varsigma^\dagger_{\beta}H^{(2)}\varsigma_{\alpha}\varsigma^\dagger_{\beta'}H^{(2)}\varsigma_{\alpha'}\right]\left({\rm tr}\left[\rho^2_{\rm in}\right]-\frac{1}{d}\right)\right)\notag\\
    &=\frac{1}{4}\sum_{(\alpha,\beta)}\sum_{(\alpha',\beta')}~\frac{1}{d^2-1}\left(
    \Delta \rho_{\rm in,1}{\rm tr}\left[\varsigma^\dagger_{\beta}H^{(2)}\varsigma_{\alpha}\right]{\rm tr}\left[\varsigma^\dagger_{\beta'}H^{(2)}\varsigma_{\alpha'}\right]
    +\Delta \rho_{\rm in,2}
    {\rm tr}\left[\varsigma^\dagger_{\beta}H^{(2)}\varsigma_{\alpha}\varsigma^\dagger_{\beta'}H^{(2)}\varsigma_{\alpha'}\right]
    \right).
\end{align}
In addition, the first term in Eq.~(\ref{apdx:U_1_haar_pre}) can be further simplified as 
\begin{align}\label{apdx:trxtr}
    \sum_{(\alpha,\beta)}\sum_{(\alpha',\beta')}{\rm tr}\left[\varsigma^\dagger_{\beta}H^{(2)}\varsigma_{\alpha}\right]{\rm tr}\left[\varsigma^\dagger_{\beta'}H^{(2)}\varsigma_{\alpha'}\right]&=\left(\sum_{(\alpha,\beta)}{\rm tr}\left[H^{(2)}\varsigma_{\alpha}\varsigma^\dagger_{\beta}\right]\right)\left(\sum_{(\alpha',\beta')}{\rm tr}\left[H^{(2)}\varsigma_{\alpha'}\varsigma^\dagger_{\beta'}\right]\right)\notag\\
    &=4\delta_{\mu\nu}{\rm tr}\left[H^{(2)}\right]\delta_{\mu'\nu'}{\rm tr}\left[H^{(2)}\right]\notag\\
    &=4\delta_{\mu\nu}\delta_{\mu'\nu'}{\rm tr}^2\left[H\right].
\end{align}
Finally, we obtain
\begin{align}
    &\mathbb{E}_{U_1,U_2}\left[S_{\mu\nu}S_{\mu'\nu'}\right]\notag\\
    &=\frac{{\rm tr}^2\left[H\right]}{d^2-1}
    \Delta \rho_{\rm in,1}\delta_{\mu\nu}\delta_{\mu'\nu'}
    +\frac{1}{4(d^2-1)}\Delta \rho_{\rm in,2}\sum_{(\alpha,\beta)}\sum_{(\alpha',\beta')}\mathbb{E}_{U_2}\left[
    {\rm tr}\left[\varsigma^\dagger_{\beta}H^{(2)}\varsigma_{\alpha}\varsigma^\dagger_{\beta'}H^{(2)}\varsigma_{\alpha'}\right]\right].
\end{align}

\end{proof}

\begin{lemma}\label{lemma:U_2_haar}
    Suppose that the random quantum circuits $U_1$ and $U_2$ are independent.
    If $U_2$ forms a unitary t-design with $t\geq2$, the expectation of the product of two matrix elements for $S$ can be evaluated as
    \begin{align}
        \mathbb{E}_{U_1,U_2}\left[S_{\mu\nu}S_{\mu'\nu'}\right]
        &=\frac{{\rm tr}^2\left[\rho_{\rm in}\right]}{d^2-1}\Delta H_1
        \delta_{\mu\nu}\delta_{\mu'\nu'}+\frac{1}{4(d^2-1)}\Delta H_2\sum_{(\alpha,\beta)}\sum_{(\alpha',\beta')}\mathbb{E}_{U_1}\left[{\rm tr}\left[\varsigma_{\alpha}\rho_{\rm in}^{(1)} \varsigma^\dagger_{\beta}\varsigma_{\alpha'}\rho_{\rm in}^{(1)} \varsigma^\dagger_{\beta'}\right]\right].
    \end{align}
    Here, the summation runs in the same way as lemma~\ref{lemma:U_1_haar}.
\end{lemma}

\begin{proof}
This proof follows the same flow in lemma~\ref{lemma:U_1_haar}.
\begin{align}
    &\mathbb{E}_{U_2}\left[S_{\mu\nu}S_{\mu'\nu'}\right]\notag\\
    &=\frac{1}{4}\sum_{(\alpha,\beta)=(\mu,\nu),(\nu,\mu)}\sum_{(\alpha',\beta')=(\mu',\nu'),(\nu',\mu')}\mathbb{E}_{U_2}~{\rm tr}\left[U_2^\dagger HU_2\varsigma_{\alpha}\rho_{\rm in}^{(1)} \varsigma^\dagger_{\beta}\right]{\rm tr}\left[U_2^\dagger HU_2\varsigma_{\alpha'}\rho_{\rm in}^{(1)} \varsigma^\dagger_{\beta'}\right]\notag\\
    &=\frac{1}{4}\sum_{(\alpha,\beta)}\sum_{(\alpha',\beta')}\sum_{\bm{i},\bm{j}}~\mathbb{E}_{U_2}~{\rm tr}\left[U_2\varsigma_{\alpha}\rho_{\rm in}^{(1)} \varsigma^\dagger_{\beta}U_2^\dagger\ket{\bm{i}}\bra{\bm{j}} HU_2\varsigma_{\alpha'}\rho_{\rm in}^{(1)}\varsigma^\dagger_{\beta'}U_2^\dagger\ket{\bm{j}}\bra{\bm{i}}H\right].
\end{align}
If $U_2$ forms a unitary $t$-design with $t\geq 2$, we obtain 
\begin{align}
    &\sum_{\bm{i},\bm{j}}~\mathbb{E}_{U_2}~{\rm tr}\left[U_2\varsigma_{\alpha}\rho_{\rm in}^{(1)} \varsigma^\dagger_{\beta}U_2^\dagger\ket{\bm{i}}\bra{\bm{j}} HU_2\varsigma_{\alpha'}\rho_{\rm in}^{(1)}\varsigma^\dagger_{\beta'}U_2^\dagger\ket{\bm{j}}\bra{\bm{i}}H\right]\notag\\
    &=\frac{1}{d^2-1}\left({\rm tr}\left[\varsigma_{\alpha}\rho_{\rm in}^{(1)} \varsigma^\dagger_{\beta}\right]{\rm tr}\left[\varsigma_{\alpha'}\rho_{\rm in}^{(1)} \varsigma^\dagger_{\beta'}\right]{\rm tr}^2\left[ H\right]+{\rm tr}\left[\varsigma_{\alpha}\rho_{\rm in}^{(1)} \varsigma^\dagger_{\beta}\varsigma_{\alpha'}\rho_{\rm in}^{(1)} \varsigma^\dagger_{\beta'}\right]{\rm tr}\left[ H^2\right]\right)\notag\\
    &~-\frac{1}{d(d^2-1)}\left({\rm tr}\left[\varsigma_{\alpha}\rho_{\rm in}^{(1)} \varsigma^\dagger_{\beta}\varsigma_{\alpha'}\rho_{\rm in}^{(1)} \varsigma^\dagger_{\beta'}\right]{\rm tr}^2\left[ H\right]+{\rm tr}\left[\varsigma_{\alpha}\rho_{\rm in}^{(1)} \varsigma^\dagger_{\beta}\right]{\rm tr}\left[\varsigma_{\alpha'}\rho_{\rm in}^{(1)} \varsigma^\dagger_{\beta'}\right]{\rm tr}\left[ H^2\right]\right),
\end{align}
where we used the formula in Eq.~(\ref{apdx:weingarten_secondmoment}). Thus, this leads to
\begin{align}\label{apdx:U_2_haar_pre}
    &\mathbb{E}_{U_2}\left[S_{\mu\nu}S_{\mu'\nu'}\right]\notag\\
    &=\frac{1}{4}\sum_{(\alpha,\beta)}\sum_{(\alpha',\beta')}~\frac{1}{d^2-1}\left(
    {\rm tr}\left[\varsigma_{\alpha}\rho_{\rm in}^{(1)} \varsigma^\dagger_{\beta}\right]{\rm tr}\left[\varsigma_{\alpha'}\rho_{\rm in}^{(1)} \varsigma^\dagger_{\beta'}\right]\left({\rm tr}^2\left[ H\right]-\frac{{\rm tr}\left[ H^2\right]}{d}\right)\right)\notag\\
    &~+\frac{1}{4}\sum_{(\alpha,\beta)}\sum_{(\alpha',\beta')}~\frac{1}{d^2-1}\left({\rm tr}\left[\varsigma_{\alpha}\rho_{\rm in}^{(1)} \varsigma^\dagger_{\beta}\varsigma_{\alpha'}\rho_{\rm in}^{(1)} \varsigma^\dagger_{\beta'}\right]\left({\rm tr}\left[ H^2\right]-\frac{{\rm tr}^2\left[ H\right]}{d}\right)\right)\notag\\
    &=\frac{1}{4}\sum_{(\alpha,\beta)}\sum_{(\alpha',\beta')}~\frac{1}{d^2-1}\left(\Delta H_1{\rm tr}\left[\varsigma_{\alpha}\rho_{\rm in}^{(1)} \varsigma^\dagger_{\beta}\right]{\rm tr}\left[\varsigma_{\alpha'}\rho_{\rm in}^{(1)} \varsigma^\dagger_{\beta'}\right]+\Delta H_2{\rm tr}\left[\varsigma_{\alpha}\rho_{\rm in}^{(1)} \varsigma^\dagger_{\beta}\varsigma_{\alpha'}\rho_{\rm in}^{(1)} \varsigma^\dagger_{\beta'}\right]\right).
\end{align}
In addition, the first term in Eq.~(\ref{apdx:U_2_haar_pre}) can be further simplified as
\begin{align}
    \sum_{(\alpha,\beta)}\sum_{(\alpha',\beta')}{\rm tr}\left[\varsigma_{\alpha}\rho_{\rm in}^{(1)} \varsigma^\dagger_{\beta}\right]{\rm tr}\left[\varsigma_{\alpha'}\rho_{\rm in}^{(1)} \varsigma^\dagger_{\beta'}\right]
    &=\left(\sum_{(\alpha,\beta)}{\rm tr}\left[\rho_{\rm in}^{(1)} \varsigma^\dagger_{\beta}\varsigma_{\alpha}\right]\right)\left(\sum_{(\alpha',\beta')}{\rm tr}\left[\rho_{\rm in}^{(1)} \varsigma^\dagger_{\beta'}\varsigma_{\alpha'}\right]\right)\notag\\
    &=4\delta_{\mu\nu}{\rm tr}\left[\rho_{\rm in}^{(1)}\right]\delta_{\mu'\nu'}{\rm tr}\left[\rho_{\rm in}^{(1)}\right]\notag\\
    &=4\delta_{\mu\nu}\delta_{\mu'\nu'}.
\end{align}
Consequently, we obtain
\begin{align}
    &\mathbb{E}_{U_1,U_2}\left[S_{\mu\nu}S_{\mu'\nu'}\right]\notag\\
    &=\frac{1}{4(d^2-1)}\mathbb{E}_{U_1}\left[\Delta H_1
    \sum_{(\alpha,\beta)}\sum_{(\alpha',\beta')}{\rm tr}\left[\varsigma_{\alpha}\rho_{\rm in}^{(1)} \varsigma^\dagger_{\beta}\right]{\rm tr}\left[\varsigma_{\alpha'}\rho_{\rm in}^{(1)} \varsigma^\dagger_{\beta'}\right]\right]\notag\\
    &~+\frac{1}{4(d^2-1)}\mathbb{E}_{U_1}\left[\Delta H_2\sum_{(\alpha,\beta)}\sum_{(\alpha',\beta')}{\rm tr}\left[\varsigma_{\alpha}\rho_{\rm in}^{(1)} \varsigma^\dagger_{\beta}\varsigma_{\alpha'}\rho_{\rm in}^{(1)} \varsigma^\dagger_{\beta'}\right]\right]\notag\\
    &=\frac{1}{d^2-1}\Delta H_1
    \delta_{\mu\nu}\delta_{\mu'\nu'}+\frac{1}{4(d^2-1)}\Delta H_2\sum_{(\alpha,\beta)}\sum_{(\alpha',\beta')}\mathbb{E}_{U_1}\left[{\rm tr}\left[\varsigma_{\alpha}\rho_{\rm in}^{(1)} \varsigma^\dagger_{\beta}\varsigma_{\alpha'}\rho_{\rm in}^{(1)} \varsigma^\dagger_{\beta'}\right]\right].
\end{align}
\end{proof}

\newpage
\renewcommand{\theequation}{d.\arabic{equation}}
\setcounter{equation}{0}
\subsection{Additional numerical experiments}

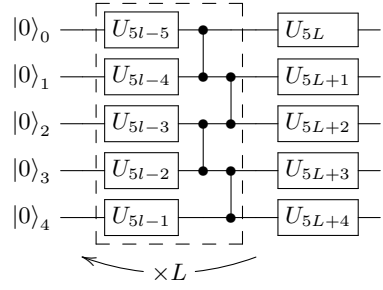
\begin{figure*}[h]
\centering
\begin{minipage}[t]{0.45\linewidth}
\begin{tabular}{c}
\Qcircuit @C=.9em @R=.45em {
  \lstickx{\ket{0}_0} & \qw & \gate{U_{5l-5}} & \ctrl{1}          & \qw              & \qw& \gate{U_{5L~~~}}   & \qw \\
  \lstickx{\ket{0}_1} & \qw & \gate{U_{5l-4}} & \control \qw & \ctrl{1}          & \qw& \gate{U_{5L+1}}       & \qw \\
  \lstickx{\ket{0}_2} & \qw & \gate{U_{5l-3}} & \ctrl{1}         & \control \qw  & \qw& \gate{U_{5L+2}}      & \qw \\
  \lstickx{\ket{0}_3} & \qw & \gate{U_{5l-2}} & \control \qw & \ctrl{1}          & \qw& \gate{U_{5L+3}}    & \qw \\
  \lstickx{\ket{0}_4} & \qw & \gate{U_{5l-1}} & \qw              & \control \qw  & \qw& \gate{U_{5L+4}}     & \qw \\
  & & & & &    \arrep{llll}
  \gategroup{1}{3}{5}{5}{.7em}{--}
}
\end{tabular}
\caption{Grouped-layer PQC with ladder entangler}
\label{fig:Grouped_PQC_ladder}
\end{minipage}
\end{figure*}

\begin{figure*}[htb]
  \centering
  \begin{tabular}{c}
  \includegraphics[scale=0.40]{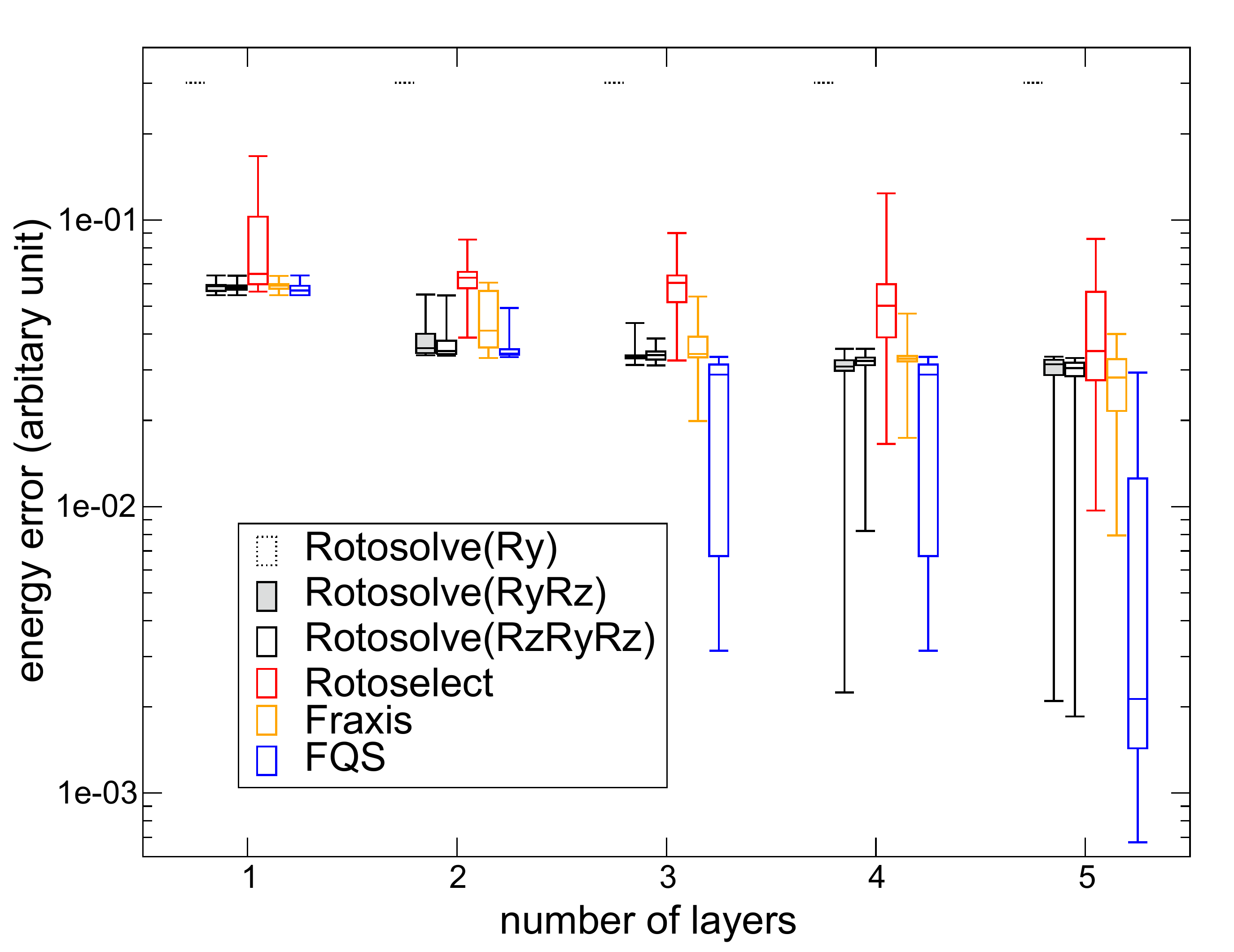}
  \end{tabular}
  \caption{Boxplots of the resulting energy error of VQE for 5-qubit 1-dimensional mixed field Ising model after 100 sweeps.
  The grouped-layer PQC with ladder entangler~\ref{fig:Grouped_PQC_ladder}, which is also an alternating layered ansatz, was employed.
  The 20 independent VQEs were conducted from randomly-generated initial states}
 \label{fig:Ising_layers_ladder}
\end{figure*}

\begin{figure*}[t]
 \centering
 \begin{tabular}{c}
\includegraphics[scale=0.40]{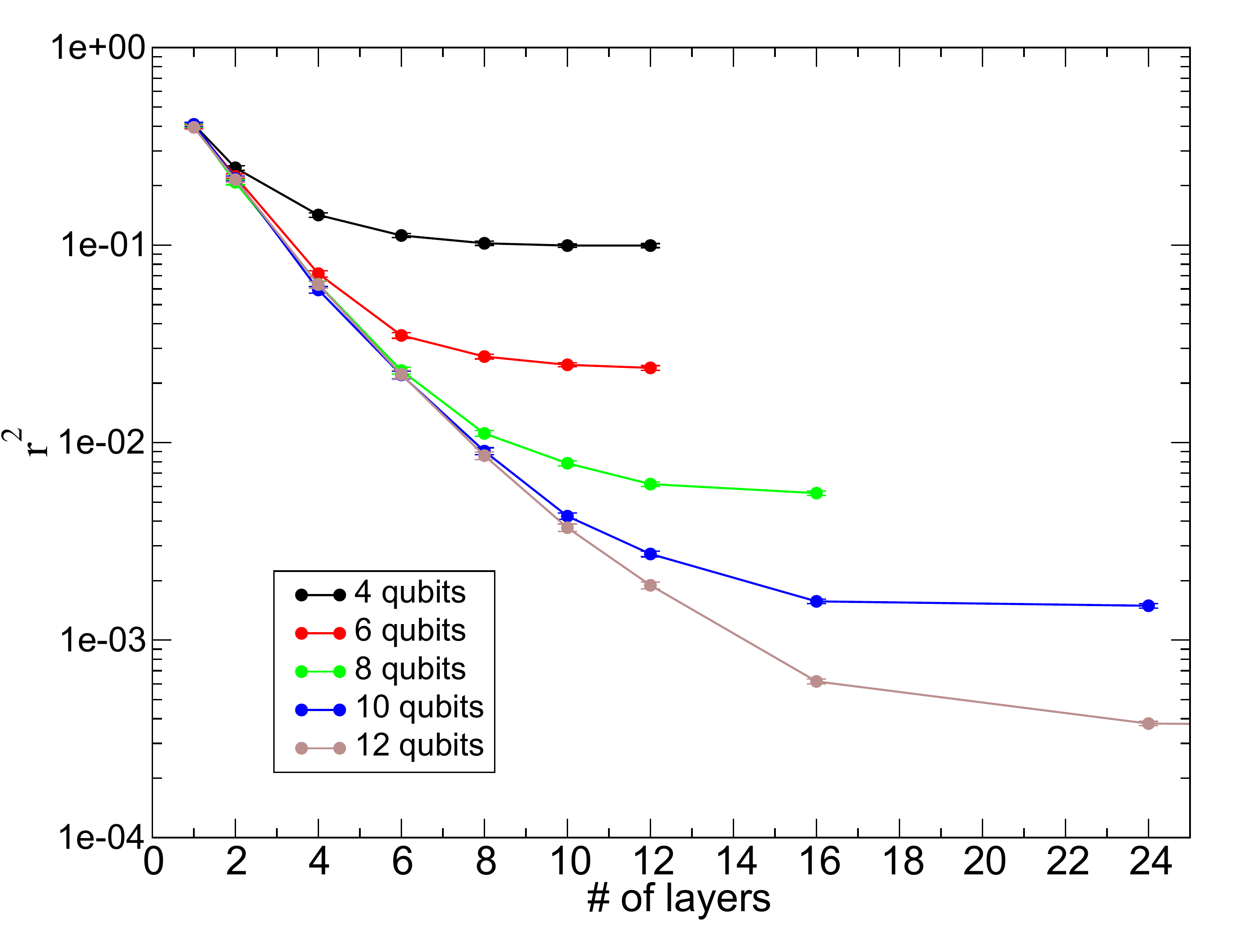} 
 \end{tabular}
 \caption{The second moment of spectrum radius of the centered FQS matrix. 
 The FQS matrix was evaluated for the single-qubit gate acting on the first qubit in the first layer of the alternating layered ansatz with ladder entanglers as shown in Fig.~\ref{fig:Grouped_PQC_ladder}.
 The second moment was evaluated with 1000 samples based on the randomly parameterized circuits.
 For the local cost function, we employed an expectation value of the Hamiltonian $H=Z\otimes I^{\otimes n-1}$. 
 }
 \label{fig:spectral_radius_ladder}
\end{figure*}

\begin{figure*}[t]
 \centering
 \begin{tabular}{c}
\includegraphics[scale=0.40]{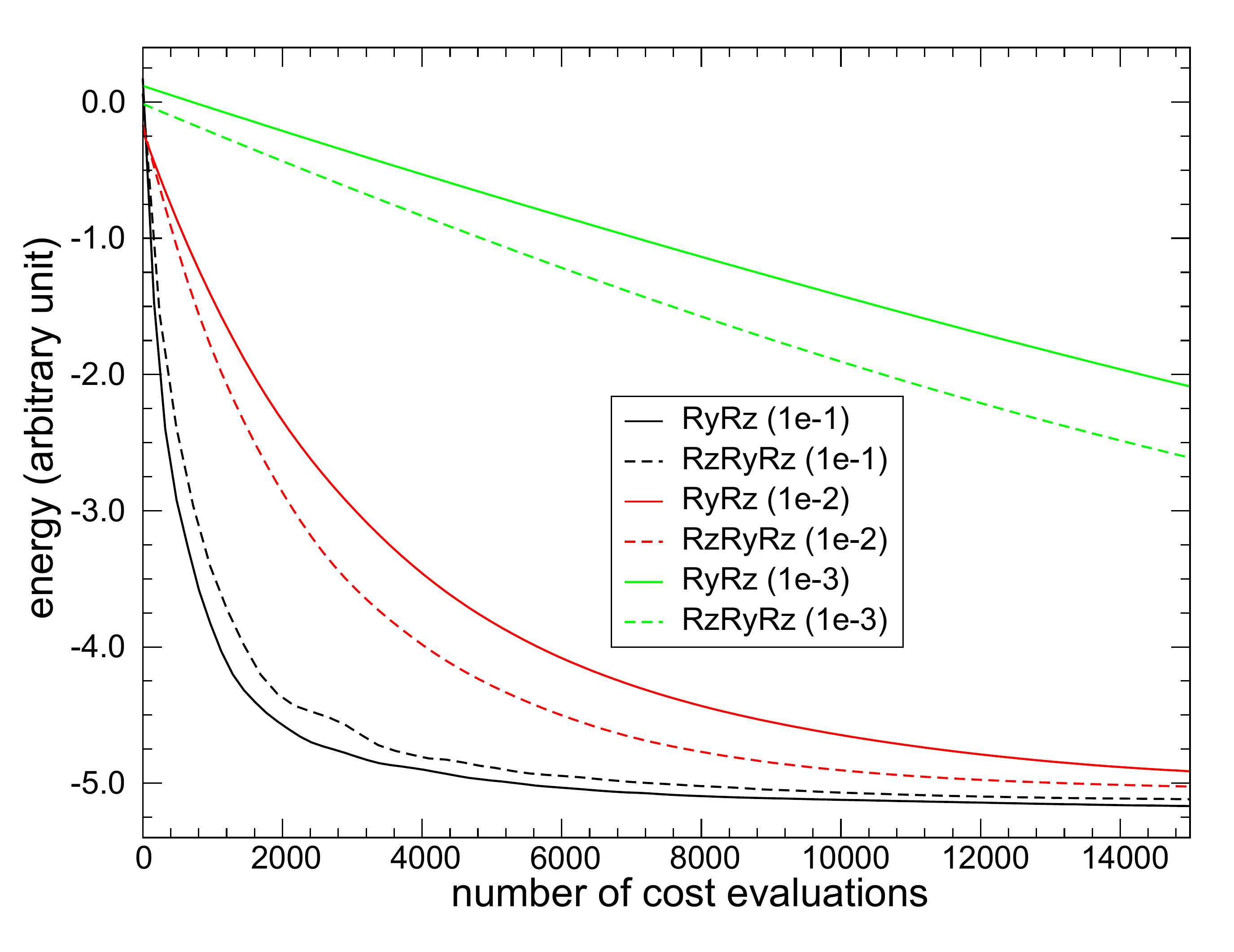} 
 \end{tabular}
 \caption{Comparison of the 
 optimization trajectories of the VQE resulting energy by the Adam optimizer with different learning rates for 5-qubit 1-dimensional mixed field Ising model. 
 Each trajectory is average over 20 independent runs.
 Black, red, and green lines represent the results of the learning rates of 0.1, 0.01 and 0.001, respectively.
 The solid and dash lines represent on the 5-layer alternating layered ansatz consisting of $RyRz$ and $RzRyRz$, respectively. 
}
 \label{fig:adam_rate}
\end{figure*}

\end{document}